\documentclass[seceq]{ptptex}

\usepackage{graphicx}
\usepackage{wrapft}

\def\CA{{\mathcal A}}

\def\CE{{\mathcal E}}

\def\CM{{\mathcal M}}

\def\CT{{\mathcal T}}

\def\bA{\mbox{\boldmath $A$}}
\def\bK{\mbox{\boldmath $K$}}
\def\bL{\mbox{\boldmath $L$}}
\def\bP{\mbox{\boldmath $P$}}

\def\bS{\mbox{\boldmath $S$}}
\def\bV{\mbox{\boldmath $V$}}

\def\bX{\mbox{\boldmath $X$}}

\def\bk{\mbox{\boldmath $k$}}
\def\bn{\mbox{\boldmath $n$}}
\def\bp{\mbox{\boldmath $p$}}
\def\bq{\mbox{\boldmath $q$}}
\def\br{\mbox{\boldmath $r$}}

\def\bx{\mbox{\boldmath $x$}}

\def\bfell{\mbox{\boldmath $\ell$}}

\def\bfsigma{\mbox{\boldmath $\sigma$}}

\def\eq#1{Eq.\,(\ref{#1})}

\def\ph{\phi^{\rm int}}

\title{
$n\alpha$ Resonating-Group Calculation with a Quark-Model
$G$-Matrix $NN$ Interaction
}

\author{
Yoshikazu \textsc{Fujiwara}$^1$, Michio \textsc{Kohno}$^2$
and Yasuyuki \textsc{Suzuki}$^3$
}

\inst{
$\hbox{}^1$ Department of Physics, Kyoto University,
Kyoto 606-8502, Japan \\
$\hbox{}^2$ Physics Division, Kyushu Dental College,
Kitakyushu 803-8580, Japan \\
$\hbox{}^3$ Department of Physics, and Graduate School
of Science and Technology,
Niigata University, Niigata 950-2181, Japan
}

\abst{
We calculate $n \alpha$ phase-shifts and scattering observables
in the resonating-group method, using the nuclear-matter $G$-matrix
of an $SU_6$ quark-model $NN$ interaction.
The $G$-matrix is generated in the recent energy-independent procedure
of the quark-model $NN$ interaction with the continuous prescription
for intermediate spectra, by assuming an appropriate Fermi momentum
$k_F=1.2~\hbox{fm}^{-1}$.
The $n\alpha$ RGM interaction kernels are evaluated
with explicit treatments of the nonlocality
and momentum dependence of partial-wave $G$-matrix components.
The momentum dependence of the $G$-matrix components is different
for each of the nucleon-exchange and interaction types.
Without introducing any artificial parameters except for $k_F$, 
the central and spin-orbit components
of the $n \alpha$ Born kernel are found to have reasonable strengths
under the assumption of a rigid translationally invariant
shell-model wave function of the $\alpha$-cluster.
The characteristic behaviors of three different exchange terms,
corresponding to knockout, heavy-particle pickup
and nucleon-rearrangement processes, are essentially the same
between the case of previous local effective $NN$ forces
and the case of nonlocal $G$-matrix $NN$ interactions.
}

\begin{document}

\maketitle

\section{Introduction}
Microscopic cluster models have been successfully used to describe
structure and reactions of light nuclear systems.
For instance, low-energy $n \alpha$ scattering is well described by
the $n \alpha$ resonating-group method (RGM) with various
model spaces and effective nucleon-nucleon ($NN$) forces. 
These effective forces usually incorporate only the central and
spin-orbit ($LS$) forces. The very strong one-pion exchange
tensor force of the bare $NN$ interaction is renormalized
to the $\hbox{}^3E$ central force effectively.
The usage of these effective forces is justified largely by the
success of the RGM calculations, in which the existence of 
ample experimental data for the $NN$ scattering and light nuclear
systems are prerequisite. On the other hand, the experimental data
in the hypernuclear systems are still not yet sufficient and basic
baryon-baryon interactions are not well known because of the 
technical difficulties of strangeness experiments.
From the theoretical side, some progress is made with the study
of baryon-baryon interactions and with accurate calculational
techniques to solve few-body systems.
One of these attempts is our effort to construct a unified set of
quark-model $B_8$-$B_8$ potentials  \cite{PPNP} and to apply them
to solve few-body systems \cite{ren}.
We now need a procedure to link bare and effective
interactions through some effective interaction theory
such as the $G$-matrix formalism.
In this paper we present such a calculational scheme for light nuclei,
in which we introduce no intermediate localized effective potential,
and directly use $G$-matrices in nuclear matter.
First, we reexamine the well-studied problem
of $n \alpha$ RGM calculations from the viewpoint of our method
to establish the reliability to proceed to hyperon-nucleus systems.
The latter applications are to be reported in a successive paper.
 
We will carry out $n \alpha$ RGM calculation using 
a quark-model $G$-matrix $NN$ interaction and a framework
that has recently been developed for $\alpha$-cluster folding
in the study of the baryon-octet ($B_8$) $\alpha$ interaction \cite{B8a}. 
In this framework, the partial-wave components of the $G$-matrix
are explicitly used to generate the direct and knock-on\footnote{We will
use in this paper the terminology ``knock-on term'' to specify
the interaction terms responsible for the ``knockout process'' in
the nuclear reaction mechanism discussed in Ref.\,\citen{TT71}.} terms
of the $n \alpha$ RGM kernel, without making any kind of
local approximation for the $G$-matrix.
The center-of-mass (c.m.) motion of two interacting nucleons
in the $n \alpha$ system is correctly treated
for the Galilean non-invariant $G$-matrix interaction.
The $G$-matrix and momentum-dependent single-particle (s.p.)~potentials
are pre-determined by solving the Bethe-Goldstone
equation in symmetric nuclear matter \cite{GMAT}.
The only assumption is a constant Fermi momentum $k_F$ for generating
the $G$-matrix interaction that can be used as an effective interaction
in light nuclear systems.
We use the energy-independent version of the quark-model $NN$ interaction
to calculate the $G$-matrix, but the difference from the previous
energy-dependent version in Ref.\ \citen{GMAT} is very little.
The original quark-model $NN$ interactions, fss2 and FSS, used in this
paper are a unified model for full octet-baryons \cite{PPNP},
which have achieved accurate descriptions
of the $NN$ and $YN$ scattering observables.
In particular, the $NN$ interaction of the most
recent model fss2~\cite{fss2} is sufficiently 
accurate in comparison with those of
modern realistic meson-exchange models.

Since the $n \alpha$ RGM kernel involves various nucleon-exchange
terms, we need to extend the previous folding formula,
starting from the transformation formula developed in
Appendix A of Ref.~\citen{LSRGM}.
In general, the interaction kernel involves five different interaction
types for two-cluster configurations with a common width parameter
of $(0s)$ clusters.
Among them, the exchange terms, called the $1S$ and $1S^\prime$ types
in this paper, do not appear in the previous hyperon-$\alpha$ interaction
and need a special treatment in the present $n \alpha$ study.
These terms correspond to the so-called heavy-particle pickup
process \cite{TT71} and play an important role in the backward increase
in the differential cross sections in the low-energy
and intermediate-energy regions.
On the other hand, the knock-on term specified by the $1D_-$ type gives
an energy-dependent extra attraction to the direct potential
specified by $0D_+$, and contributes mainly to the forward direction
in the usual RGM treatment with effective $NN$ forces.
In the present formalism,
all these interaction kernels have their own momentum dependence
for the Galilean non-invariant two-nucleon interaction.
We can calculate the momentum dependence explicitly in the analytic
form. A very important starting-energy dependence of the $G$-matrix
is renormalized into this momentum dependence and the dependence 
to the relative momentum, and to the Fermi momentum $k_F$ as well.
Owing to the explicit evaluation of all the interaction kernels,
the existence of the Pauli forbidden $(0s)$ state between the neutron
and the $\alpha$ cluster is strictly preserved.
We will find that the present procedure gives rather reasonable strengths
of the $n \alpha$ interaction for both the central
and $LS$ components, and reproduces reasonably well the empirical
$S$-wave and $P$-wave phase shifts and the low-energy
differential cross sections and polarizations below
the neutron incident energy $E_n \sim 30$ MeV.
For higher energies, we investigate the $n \alpha$ Born kernel and
find that the characteristic behavior of three different groups
of exchange terms, corresponding to knockout, heavy-particle pickup
and nucleon-rearrangement processes,
found in previous studies using local effective $NN$ forces \cite{TT71},
is essentially unchanged even in the present study
with the nonlocal $G$-matrix interaction.

The $n \alpha$ RGM has been examined by many authors from many
different viewpoints. In fact, early studies have naturally
paid full attention to the adequacy of the model space
and the $NN$ force used in the calculation. For example,
Sugie {\em et al.} \cite{SU55} studied the contribution
of tensor force, trying to explain the large energy splitting
of the $J^\pi=3/2^-$ and $1/2^-$ states. Since the inherent $LS$ force
of the $NN$ interaction has been found after this calculation, the authors
did not include it and found that tensor force can account for
only about 30\% of the observed splitting. Later calculations by
Kanada {\em et al.} \cite{KA63} and Omojola \cite{OM70} included
the $LS$ force and have found that it is important to deal with
the $D$-state components of the $\alpha$ cluster
and realistic $NN$ interactions.
In these calculations, an approximate Hamada and Johnston potential
is used in the Gaussian form. They reproduced the low-energy $n\alpha$
phase shifts reasonably well. The channel coupling effect
of the $n \alpha$ and $d\,\hbox{}^3\hbox{H}$ configurations are examined 
by Heiss and Hackenbloich \cite{HE70}.
Thompson and Tang \cite{TT71} analyzed the properties of the $n \alpha$
RGM exchange kernels for an effective central $NN$ force,
and classified them into three different groups
of terms, called the knockout, heavy-particle pickup
and nucleon-rearrangement terms.
Chwieroth {\em et al.} \cite{CH74} and later
calculations \cite{KA74,SH85} have clarified that the distortion effect
of the $\alpha$ cluster also has an appreciable effect even in
single-channel $n \alpha$ RGM calculation. 
Nevertheless, the effect is rather minor, in comparison with the other
two-cluster systems, since the $\alpha$ particle is tightly bound.
After all of these investigations, it is gradually recognized that
single-channel $n \alpha$ RGM calculation, using a rigid $(0s)^4$
$\alpha$-cluster wave function and a simple effective $NN$ force of the
central and $LS$ types, reproduces the low-energy $n \alpha$ scattering
fairly well, except for some specific energy regions
where other reaction channels open.
This understanding cannot directly be reconciled with the recent
{\em ab~initio} calculation by Nollett {\em et al.} \cite{NO07},
who claim that the correct $P$-wave spin-orbit splitting
of the $n \alpha$ scattering phase shifts in the low-energy
region can only be achieved with the effect of an appropriate
three-nucleon interaction.

The organization of this paper is as follows.
In the next section, we first recapitulate in Sec.~2.1 the standard
RGM formulation in the momentum representation, together with
the $G$-matrix calculation of the quark-model baryon-baryon
interaction for symmetric nuclear matter.
The full expressions of the $n \alpha$ exchange interaction kernels
for the $G$-matrix $NN$ interaction are given in Sec.~2.2.
The partial-wave decomposition is presented in Sec.~2.3.
In Sec.~2.4, we discuss the selection of the starting energy
in the $G$-matrix calculation.
Numerical results of the $G$-matrix calculation
and the $n \alpha$ RGM phase shifts are given in Sec.~3.1.
The $n \alpha$ scattering cross sections and polarization
for neutron incident energies less than 30 MeV
are compared with experimental results in Sec.~3.2.
The characteristic behavior of the $n \alpha$ Born amplitudes
for higher energies is analyzed in Sec.~3.3.
Section 4 is devoted to the summary of this paper.
The $n \alpha$ RGM Born kernels for a Gaussian-type effective $NN$ force
are given in Appendix A.
In Appendix B, we give the exchange interaction kernels of a Galilean
non-invariant $NN$ interaction for general systems,
composed of two $(0s)$-shell clusters.

\section{Formulation}

\vspace{-3mm}

\subsection{Lippmann-Schwinger RGM and the $G$-matrix calculation
of symmetric nuclear matter}


For the correct treatment of the c.m. motion, it is most convenient
to formulate the RGM in the momentum representation,
which we call the Lippmann-Schwinger RGM (LS-RGM) \cite{LSRGM}.
In this approach, we write the RGM equation in the Schr{\"o}dinger-type
equation
\begin{eqnarray}
& & \left[\varepsilon - T_r-V^{\rm RGM}(\varepsilon)\right] \chi =0 \ ,
\label{rgm1}
\end{eqnarray}
where $T_r$ is the kinetic-energy operator for the relative motion
of two clusters, $\varepsilon=E-E_{\rm int}$ with $E_{\rm int}$ being
the internal energy, and $V^{\rm RGM}(\varepsilon)=V_{\rm D}
+G+\varepsilon K$ the sum of the direct potential $V_{\rm D}$,
the exchange kernel $G$, and the energy-dependent term $\varepsilon K$,
inherent in the RGM formalism.
We use the notation $K$ for the exchange normalization kernel,
which is defined through
\begin{equation}
N=\langle \ph \vert \CA^\prime \vert \ph \rangle=1+X_N M_N=1-K\ .
\label{rgm2}
\end{equation}
In the $n\alpha$ RGM, the spin-isospin factor
for the exchange normalization kernel is $X_N=-1$,
and $M_N$ is the corresponding spatial part.
The total wave function is expressed as $\Psi=\CA^\prime \{ \ph \chi \}$
using the relative wave function $\chi$ in \eq{rgm1}.
Here, $\ph$ is the internal wave function and $\CA^\prime$ the
antisymmetrization operator between two clusters.
For the $n\alpha$ system, these are given by
$\ph=\phi_\alpha~\xi_{\frac{1}{2}\frac{1}{2}}$ and  
$\CA^\prime=1-\sum^4_{i=1}P_{(i5)} \rightarrow 1-4 P_{(45)}$
using the internal wave function of the $\alpha$ cluster
$\phi_\alpha=\phi^{\rm orb}_\alpha~\xi_{00}$
with $\phi^{\rm orb}_\alpha$ being the spatial part of the translationally
invariant $(0s)^4$ harmonic-oscillator (h.o.) wave function
without the c.m.~motion. The spin-isospin wave functions are denoted
by $\xi_{00}$ and $\xi_{\frac{1}{2}\frac{1}{2}}$ for
the $\alpha$ cluster and the fifth nucleon, respectively.
In this particular case, we have one Pauli-forbidden
state $\vert u_{(0s)} \rangle$ satisfying
\begin{equation}
K \vert u_{(0s)} \rangle=\vert u_{(0s)} \rangle \qquad \hbox{and} \qquad
\CA^\prime \{ \ph u_{(0s)} \}=0\ .
\label{rgm3}
\end{equation}
The exchange kernel $G=G^{\rm K}+G^{\rm V}$ is composed of the
exchange kinetic-energy kernel $G^{\rm K}$ and the exchange
interaction kernel $G^{\rm V}$ for the total Hamiltonian consisting of 
\begin{equation}
H=\sum^5_{i=1} t_i-T_G+\sum^5_{i<j} v_{ij}
\label{rgm4}
\end{equation}
in the $n \alpha$ RGM.
Equation (\ref{rgm1}) is then solved in the
form of the Lippmann-Schwinger equation for
the half-off shell $T$-matrix with discretized momentum
mesh points, which is fully spelled out in Ref.~\citen{LSRGM}.

The main task in the LS-RGM is therefore to calculate the plane-wave
matrix elements of various exchange kernels, which we call
the RGM Born kernels.
For the interaction kernels, we separate the two-nucleon force
into the spatial and spin-isospin parts,
using $v_{ij}=u_{ij}\,w_{ij}$. The direct potential $V_{\rm D}$ and the exchange
interaction kernel $G^{\rm V}$ are obtained from the general expression
\begin{eqnarray}
M(\bq_f, \bq_i) & = & \langle\,\delta(\bX_G)\,e^{i \bq_f \cdot \br}~\ph\,\vert
\sum_{i<j} v_{ij} \CA^\prime \,\vert
\,1 \cdot e^{i \bq_{\,i}\cdot \br}~\ph \rangle \nonumber \\
& = & \sum_{x \CT} X_{x \CT}
~M_{x \CT}(\bq_f, \bq_i)\ \ .
\label{rgm5}
\end{eqnarray}
Here, the spatial integral $M_{x \CT}(\bq_f, \bq_i)$ and the spin-isospin
factor $X_{x \CT}$ are defined by
\begin{eqnarray}
M_{x \CT}(\bq_f, \bq_i)
& = & \langle\,\delta(\bX_G)\,z_x~e^{i \bq_f \cdot \br}\,\phi^{\rm orb}
\,\vert\,u_{ij}\,\vert 1\cdot e^{i \bq_{\,i}\cdot \br}
\,\phi^{\rm orb} \rangle\ ,\nonumber \\
X_{x \CT} & = & C_x \langle z_x~\xi\,|\,\sum_{i<j}^\CT w_{ij}
\,|\,\xi \rangle \quad \hbox{with} \quad (i,j) \in \CT\ . 
\label{rgm6}
\end{eqnarray}
In the $n\alpha$ RGM, $\phi^{\rm orb}=\phi^{\rm orb}_\alpha$ and
$\xi=\xi_{00}\,\xi_{\frac{1}{2}\frac{1}{2}}$.
These kernel components are specified by
the number of exchanged nucleons, $x=0,~1$, with $C_0=0$, $C_1=-4$,
and various interaction types $\CT$.
More specifically, $x=0$ with $z_0=1$ corresponds
to the direct terms including $\CT=E$ and $D_+$ interaction types
and $x=1$ with $z_1=P_{(45)}$ the one-nucleon exchange terms.
The interaction types, $\CT=E,~S,~S^\prime,~D_+$ and $D_-$, correspond
to some specific $(i,j)$ pairs of the nucleons \cite{KI94}.
When an effective $NN$ force is used, these Born kernels are most
easily calculated using a general transformation formula developed
in Appendix A of Ref.\,\citen{LSRGM}.
For Gaussian-type effective $NN$ forces with
the central form
\begin{eqnarray}
v^{(C)}=u(r)\,w=v_0\,e^{-\kappa r^2}\,\left(W+BP_\sigma-HP_\tau
-MP_\sigma P_\tau\right)\ ,
\label{rgm7}
\end{eqnarray}
and the $LS$ form
\begin{eqnarray}
v^{(LS)}=u^{LS}(r)\,w^{LS}=v^{LS}_0\,e^{-\kappa r^2}
\,\left(W-HP_\tau\right)\,(\bL \cdot \bS)\ ,
\label{rgm8}
\end{eqnarray}
the final results of the $n\alpha$ Born kernels are given
in Appendix A for completeness.

The quark-model baryon-baryon interaction is formulated in 
a similar way to the $n \alpha$ RGM \cite{PPNP}. In this case, the
internal wave function is a product of two three-quark
clusters $(3q)$-$(3q)$. We solve the $G$-matrix equation
\begin{eqnarray}
& &  G_{NN}(\bp, \bq; K, \omega, k_F)
=V^{\rm RGM}_{NN}(\bp, \bq) \nonumber \\
& & \hspace{10mm} +\frac{1}{(2\pi)^3} \int d~\bk
~V^{\rm RGM}_{NN}(\bp, \bk)
\frac{Q(k, K, k_F)}{e(k, K; \omega)}
~G_{NN}(\bk, \bq; K, \omega, k_F)\ ,
\label{rgm9}
\end{eqnarray}
using the energy-independent Born kernel
$V^{\rm RGM}_{NN}(\bp, \bq)$ for the $(3q)$-$(3q)$ system \cite{GMAT}.
Here, $V^{\rm RGM}_{NN}(\bp, \bq)$ is defined by \cite{RRGM}
\begin{eqnarray}
V^{\rm RGM}_{NN}(\bp, \bq)
=V_{\rm D}(\bp, \bq)+G(\bp, \bq)+W(\bp, \bq)\ ,
\label{rgm9-1}
\end{eqnarray}
with
\begin{eqnarray}
W=\frac{1}{\sqrt{N}}\left(T_r+V_{\rm D}+G\right)\frac{1}{\sqrt{N}}
-\left(T_r+V_{\rm D}+G\right)\ .
\label{rgm9-2}
\end{eqnarray}
Details of this energy-independent treatment of the
quark-model baryon-baryon interaction in the $G$-matrix
formalism will be published elsewhere.
In \eq{rgm9}, we use the angle-averaged Pauli operator $Q(k, K, k_F)$,
and the energy denominator $e(k, K; \omega)$ is given by
\begin{eqnarray}
e(k, K; \omega)=\omega-E_N(k_1)-E_N(k_2),
\label{rgm10}
\end{eqnarray}
with the starting energy $\omega$ expressed as
\begin{eqnarray}
\omega=E_N(q_1)+E_N(q_2).
\label{rgm11}
\end{eqnarray}
The s.p.~momenta $\bq_1$ and $\bq_2$ ($\bk_1$ and $\bk_2$) are related
to the relative momentum $\bq$ ($\bk$) and the c.m.~momentum $\bK$
through the conventional relationship
$\bq=(\bq_1-\bq_2)/2$ and $\bK=\bq_1+\bq_2$
($\bk=(\bk_1-\bk_2)/2$ and $\bK=\bk_1+\bk_2$).
The nucleon s.p.~potential $U_N(q_1)$
included in the s.p.~energy
\begin{eqnarray}
E_N(q_1)=\frac{\hbar^2}{2M_N}{q_1}^2+U_N(q_1)
\label{rgm12}
\end{eqnarray}
is determined self-consistently in the standard procedure.
In \eq{rgm12}, $M_N$ is the nucleon mass. The procedure to include
the s.p.~potential even in the intermediate spectra in \eq{rgm10} is
called a continuous prescription.

\subsection{Interaction kernels
for $G$-matrix $NN$ interaction}


In this subsection, we derive the interaction Born kernel
for the $G$-matrix $NN$ interaction.
The starting point is the invariant $G$-matrix \cite{B8a} expressed
as\footnote{The invariant $NN$ $G$-matrix in \eq{int1} is defined 
without the factor 2, shown in Eq.\,(2.4) of Ref.~\citen{B8a},
since the exchange terms are explicitly calculated in the present
$n \alpha$ RGM formalism.} 
\begin{eqnarray}
& & G^{I}_{NN}(\bp, \bp^\prime; K, \omega, k_F) \nonumber \\
& & =\frac{1}{2} \langle\,[NN]_{II_z}\,|\,G(\bp, \bp^\prime; K, \omega,k_F)
- G(\bp, -\bp^\prime; K, \omega, k_F)\,P_\sigma\,P_\tau\,|
\,[NN]_{II_z}\,\rangle\ \ \nonumber \\
& & = g^I_0+g^I_{ss} (\bfsigma_1 \cdot \bfsigma_2)
+h^I_0\,i \widehat{\bn} \cdot (\bfsigma_1 + \bfsigma_2)
+ \cdots\ .
\label{int1}
\end{eqnarray} 
Here $\widehat{\bn}=[\bp^\prime \times \bp]/(p^\prime p \sin \theta)$,
and the invariant functions $g^I_0$ (central), $g^I_{ss}$ (spin-spin),
$h^I_0$ ($LS$), etc.~are functions of $p=|\bp|$, $p^\prime=|\bp^\prime|$,
and $\cos \theta=(\widehat{\bp} \cdot \widehat{\bp}^\prime)$,
as well as the $G$-matrix parameters $K$, $\omega$ and $k_F$.
These are expressed by the partial-wave components
of the $NN$ $G$-matrix as in Appendix D of Ref.~\citen{LSRGM}.
In the following, we will focus on the momentum dependence 
of the $G$-matrix, and keep only the parameter $K$ 
in $G^{I}_{NN}(\bp, \bp^\prime; K, \omega, k_F)$,
since the explicit dependence depends on the interaction type.
As in Ref.~\citen{B8a}, it is convenient to write
the isospin dependence of the invariant $G$-matrix as
\begin{equation}
G_{NN}(\bp, \bp^\prime; K)
=G^{I=1}_{NN}(\bp, \bp^\prime; K) \frac{1+P_\tau}{2}
+G^{I=0}_{NN}(\bp, \bp^\prime; K) \frac{1-P_\tau}{2}\ \ .
\label{int3}
\end{equation}
We find it convenient to separate the isospin multiplicity
factor $(2I+1)$ and define the spin-isospin factors
in \eq{rgm6} by
\begin{equation}
X^\Omega_{x\CT}=(2I+1)\,X^{\Omega\,I}_{x\CT}\ .
\label{int4}
\end{equation}
The interaction species $\Omega=0,~ss$ and $LS$, correspond to
the invariant functions, $g_0$, $g_{ss}$ and $h_0$, respectively.     
Table 1 lists $X^{\Omega\,I}_{x\CT}$ for each of these species.

\begin{table}[t]
\caption{Spin-isospin factors $X^{\Omega\,I}_{x\CT}$
for the invariant species, $\Omega=0$ ($g_0$), $ss$ ($g_{ss}$)
and $LS$ ($h_0$), in \protect\eq{int4}.
The factors for $x\CT=1E$, $1S$ and $1S^\prime$ are obtained from
$X^{\Omega\,I}_{1E}=X^{\Omega\,I}_{1S}=X^{\Omega\,I}_{1S^\prime}
=(-1/2) X^{\Omega\,I}_{0E}$. For the $LS$ term, $\bX^{LS\,I}_{x\CT}
=X^{LS\,I}_{x\CT}\cdot 2 \bS$ is assumed with the neutron
spin operator $\bS$.
}
\vspace{2mm}
\label{table1}
\begin{center}
\renewcommand{\arraystretch}{1.2}
\setlength{\tabcolsep}{5mm}
\begin{tabular}{ccrrr}
\hline
$I$ & $\Omega$ & $X^{\Omega\,I}_{0E}$ & $X^{\Omega\,I}_{0D_+}$
 & $X^{\Omega\,I}_{1D_-}$ \\
\hline 
  &  0   &    1 & 1 & $-\frac{1}{2}$ \\
1 & $ss$ & $-3$ & 0 & $-\frac{3}{2}$ \\
  & $LS$ & $-$  & 1 & $-1$ \\
\hline
  &  0   &    3 & 1 & $\frac{1}{2}$ \\
0 & $ss$ &    3 & 0 & $\frac{3}{2}$ \\
  & $LS$ &  $-$ & 1 &    1 \\
\hline
\end{tabular}
\end{center}
\end{table}

The spatial integrals are obtained by assuming a general
Galilean non-invariant interaction
\begin{equation}
\langle \bp_1, \bp_2 | u | \bp^\prime_1, \bp^\prime_2 \rangle
=\delta(\bK-\bK^\prime)\frac{1}{(2\pi)^3}
u(\bk^\prime, \bq^\prime; |\bK|)\ \ ,
\label{int5}
\end{equation}
where $u$ is $g^I_0$, $g^I_{ss}$, or $h^I_0\,i \widehat{\bn}$.
Here the relative momentum $\bp$ and the
total momentum $\bK$ (and also $\bp^\prime$ etc. with primes)
are related to $\bp_1$ and $\bp_2$ by
$\bp=(\bp_1-\bp_2)/2$ and $\bK=\bp_1+\bp_2$, respectively,
and a further transformation from $\bp$ and $\bp^\prime$ to
$\bk^\prime=\bp-\bp^\prime$ and $\bq^\prime=(\bp+\bp^\prime)/2$ is
applied.
The necessary spatial integrals for the $n \alpha$ system
is obtained from more general expressions
given in Appendix B for systems of two $(0s)$-shell clusters.

For the Galilean non-invariant $G$-matrix interaction,
the subtraction of the internal-energy part of the $\alpha$ cluster
involves a subtle problem. 
In the total c.m.~system of the $n \alpha$ system, the $0E$-type
spatial integral, involving the $\alpha$-cluster internal-energy
contribution, becomes momentum-dependent.
Similarly, the $1E$-exchange-type spatial integral
also involves a momentum dependence besides
the exchange normalization kernel $M_N(\bq_f, \bq_i)$. These are
explicitly given by
\begin{eqnarray}
\ \hspace{-8mm} M_{0E}(\bq_f, \bq_i)=(2\pi)^3 \delta(\bk)\,E^V_{0E}(q)\ ,
\quad M_{1E}(\bq_f, \bq_i)=M_N(\bq_f, \bq_i)\,E^V_{1E}(q)\ ,
\label{int7}
\end{eqnarray}
where $\bk=\bq_f-\bq_i$, $\bq=(\bq_f+\bq_i)/2$,
and $E^V_{xE}(q)$ with $x=0,~1$ are calculated from
\begin{eqnarray}
E^V_{0E}(q) & = & \left(\frac{1}{2\pi \nu}\right)^{\frac{3}{2}}
\int \bK\,e^{-\frac{1}{2\nu}\left(\bK-\frac{1}{2}\bq\right)^2}
\,\CE(K)\ ,\nonumber \\
E^V_{1E}(q) & = & \left(\frac{3}{4\pi \nu}\right)^{\frac{3}{2}}
\int \bK\,e^{-\frac{3}{4\nu}\left(\bK-\frac{4}{3}\bq\right)^2}
\,\CE(K)\ ,
\label{int8}
\end{eqnarray}
with
\begin{eqnarray}
\CE(K)=\frac{1}{(2\pi)^3} \left(\frac{1}{\pi \nu}\right)^{\frac{3}{2}}
\int d \bp\,d \bp^\prime~\exp \left\{-\frac{1}{2\nu}
\left(\bp^2+{\bp^\prime}^2 \right)\right\}
~g(\bp, \bp^\prime; K)\ .
\label{int9}
\end{eqnarray}
Here $g(\bp, \bp^\prime; K)=u(\bk^\prime, \bq^\prime; K)$ and
$\nu$ is the h.o.~width parameter for the $(0s)^4$ $\alpha$ cluster.
Owing to the different $q$ dependence involved in the $0E$ and $1E$ types,
the standard subtraction of the $\alpha$-cluster internal energy
in $V^{\rm RGM}(\varepsilon)$ is not complete and is given by
\begin{eqnarray}
V^{\rm RGM}(\varepsilon) & = & V_{\rm D}+G^{\rm K}+\widetilde{G}^{\rm V}
+\varepsilon K\ ,\nonumber \\ 
V_D & = & X_{0D_+} M_{0D_+}\ ,\nonumber \\
\widetilde{G}^V & = & \frac{1}{2}X_{0E} \left(
\widetilde{M}_{1E}-M_{1S}-M_{1S^\prime}\right)+X_{1D_-} M_{1D_-}
\ ,\nonumber \\
\widetilde{M}_{1E} & = & 2 M_{0E}-M_{1E}-2 M^{(0)}_{0E} (1-K)\ .
\label{int10}
\end{eqnarray}
Here we have omitted $\Omega$ and the isospin sum for simplicity. 
In deriving \eq{int10}, we have assumed that
only the $q=0$ part of $E^V_{0E}(q)$ contributes
to the $\alpha$ internal energy and defined the relative
energy $\varepsilon$ using this zero-momentum $\alpha$ energy.
In \eq{int10}, $M^{(0)}_{0E}$ implies
\begin{eqnarray}
M^{(0)}_{0E}(\bq_f, \bq_i)=(2\pi)^3 \delta(\bk)\,E^V_{0E}(0)\ .
\label{int11}
\end{eqnarray}
Since the delta function part in $\widetilde{M}_{1E}$ of \eq{int10}
is inconvenient for the exchange interaction kernel, we assume
\begin{eqnarray}
\widetilde{M}_{1E}(\bq_f, \bq_i)
=M_N(\bq_f, \bq_i)\,\left[2E^V_{0E}(q)-E^V_{1E}(q)\right]\ ,
\label{int12}
\end{eqnarray}
without violating the redundancy property of the interaction kernel.
In fact, the $q$ dependence in $E^V_{0E}(q)$ and $E^V_{0E}(q)$
is very weak for the wide range of $q$ and the difference
between $E^V_{0E}(q)$ and $E^V_{1E}(q)$ is also very small.

Some simplification takes place in the RGM calculation
using partial-wave components of the invariant interaction.
We can easily show that for each isospin component 
the direct term ($0D_+$ type) and the knock-on term ($1D_-$ type) give
the same contribution, when the central ($\Omega=0$) and
spin-spin ($\Omega=ss$) components are added up.
For the central $1E$-, $1S$- and $1S^\prime$-type spatial integrals,
only the $S$-wave component of the two-nucleon interaction contributes
to the exchange kernel.
Thus, by incorporating the spin-isospin factors given
in Table \ref{table1}, we find for the central $n \alpha$ RGM kernel 
(we remove the tilde of $\widetilde{M}_{1E}$ using
the approximation in \eq{int12})
\begin{eqnarray}
& & V^{\rm RGM}(\varepsilon)
=(V_{\rm D}+G^{\rm V})+G^{\rm K}+\varepsilon K\ ,\nonumber \\
& & V_{\rm D}+G^{\rm V}=2 \sum^1_{I=0} (2I+1)
\left(M^{0I}_{0D_+}+M^{0I}_{1E}-M^{0I}_{1S}-M^{0I}_{1S^\prime}\right)\ ,
\label{int13}
\end{eqnarray}
where $\Omega I=0 I$ spatial functions are
calculated from the $g^I_0$ (central) component in \eq{int1}. 
Similarly, the $LS$ RGM kernel is obtained only from the $0D_+$-type
contribution using the $h^I_0$ component in \eq{int1}.
The internal energy of the $\alpha$ cluster is given by
\begin{eqnarray}
E_\alpha(q)=3 \cdot \frac{3\hbar^2\nu}{2M}
+4\left[ 3 E^{V\,01}_{0E}(q)+ E^{V\,00}_{0E}(q)\right]
+2 e^2 \sqrt{\frac{\nu}{\pi}}\ ,
\label{int14}
\end{eqnarray}
where $E^{V\,0I}_{0E}(q)$ is the central component of $E^V_{0E}(q)$
with an isospin $I=0,~1$. 

To calculate the RGM kernel,
it is sometimes convenient to use either of the forms
$g(\bp, \bp^\prime; K)$ and $u(\bk^\prime, \bq^\prime; K)$,
which are equal to each other
with $\bk^\prime=\bp-\bp^\prime$ and $\bq^\prime=(1/2)(\bp+\bp^\prime)$.
The kernel expressions become sometimes simple,
if we use $\bk=\bq_f-\bq_i$ and $\bq=(1/2)(\bq_f+\bq_i)$ instead
of $\bq_f$ and $\bq_i$.
For example, the exchange normalization kernel
$M_N(\bq_f, \bq_i)=K(\bq_f, \bq_i)$ and the exchange kinetic-energy kernel
$G^{\rm K}(\bq_f, \bq_i)$ are given in Eqs.\,(\ref{a1}) and (\ref{a2}).
The $E$-type integrals are given in Eqs.\,(\ref{int7}) - (\ref{int9}).
For the direct term, we obtain
\begin{eqnarray}
M_{0D_+}(\bq_f, \bq_i) & = & e^{-\frac{3}{32\nu}\bk^2}
~\left(\frac{8}{3\pi \nu}\right)^{\frac{3}{2}}
\int d \bq^\prime~\exp \left\{-\frac{8}{3\nu}
\left(\bq^\prime-\frac{5}{8}\bq \right)^2\right\}
\nonumber \\ [2mm]
& & \times u(\bk, \bq^\prime; 2|\bq-\bq^\prime|)\ ,
\label{int15}
\end{eqnarray}
and the $S$- and $S^\prime$-type integrals are given by
\begin{eqnarray}
& & M_{\scriptsize \left\{\begin{array}{c}
1S \\
1S^\prime \\
\end{array}\right\}}(\bq_f, \bq_i)=M_N(\bq_f, \bq_i)
~\frac{1}{(2\pi)^3} \left(\frac{3}{\pi \nu}\right)^{\frac{3}{2}}
\nonumber \\ [0mm]
& & \qquad \qquad \times
\int d \bp\,\int d \bp^\prime
\,\exp \left\{ \begin{array}{c}
-\frac{1}{2\nu}(5 \bp^2+{\bp^\prime}^2)
+\frac{4}{\nu}\bp \left(\bq_f+\frac{1}{4}\bq_i\right)
-\frac{4}{3\nu}\left(\bq_f+\frac{1}{4}\bq_i\right)^2 \\
-\frac{1}{2\nu}(\bp^2+5{\bp^\prime}^2)
+\frac{4}{\nu}\bp^\prime \left( \frac{1}{4}\bq_f+\bq_i\right)
-\frac{4}{3\nu}\left(\frac{1}{4}\bq_f+\bq_i\right)^2 \\
\end{array} \right\} \nonumber \\ [2mm]
& & \qquad \qquad \times \left\{
\begin{array}{c}
g(\bp, \bp^\prime; 2|\bq_f-\bp|) \\
g(\bp, \bp^\prime; 2|\bq_i-\bp^\prime|) \\
\end{array} \right\}\ .
\label{int16}
\end{eqnarray}

\subsection{Partial-wave expansion}

The partial-wave decomposition of the $n \alpha$ RGM kernels
derived in the preceding section can be carried out
for two different types of the partial-wave decomposition
of the invariant $G$-matrix $NN$ interaction:
\begin{eqnarray}
& & g(\bp, \bp^\prime; K)=\sum^\infty_{\ell=0}
(2 \ell+1)\,g_\ell(p, p^\prime; K)
\,P_\ell(\widehat{\bp}\cdot \widehat{\bp}^\prime)\ ,\nonumber \\
& &  u(\bk^\prime, \bq^\prime; K)=\sum^\infty_{\lambda=0}
(2 \lambda+1)\,u_\lambda(k^\prime, q^\prime; K)
\,P_\lambda(\widehat{\bk}^\prime\cdot \widehat{\bq}^\prime)\ ,
\label{par1}
\end{eqnarray}
for the central part and
\begin{eqnarray}
& & h(\bp, \bp^\prime; K)=\sum^\infty_{\ell=1}
(2 \ell+1)\,h_\ell(p, p^\prime; K)
\,P^1_\ell(\widehat{\bp}\cdot \widehat{\bp}^\prime)\ ,\nonumber \\
& &  u^{LS}(\bk^\prime, \bq^\prime; K)=\sum^\infty_{\lambda=1}
(2 \lambda+1)\,u^{LS}_\lambda(k^\prime, q^\prime; K)
\,P^1_\lambda(\widehat{\bk}^\prime\cdot \widehat{\bq}^\prime)\ ,
\label{par2}
\end{eqnarray}
for the $LS$ part.
Here, the $u$-type partial-wave components are calculated from
\begin{eqnarray}
& & \ \hspace{-5mm} \left. \begin{array}{l}
g^I_{0\,\ell}(p, p^\prime; K) \\ [2mm]
g^I_{ss\,\ell}(p, p^\prime; K) \\
\end{array} \right\}
=\frac{1}{4} \sum^\prime_{JS}
\left(\frac{2J+1}{2\ell+1}\right)
\left\{\begin{array}{c}
1 \\ [2mm]
\frac{1}{3}[2S(S+1)-3] \\
\end{array} \right\}
G^{IJ}_{S\ell, S\ell}(p, p^\prime; K)
\nonumber \\
& & \hspace{100mm} (\ell=0,~1,~2,\cdots)\ ,\nonumber \\
& & \ \hspace{-5mm} h^I_{0\,\ell}(p, p^\prime, K)
=-\frac{1}{4}\left[
\frac{1}{\ell(\ell+1)} G^{I\,\ell}_{1\ell, 1\ell}(p, p^\prime; K)
+\frac{2\ell-1}{\ell(2\ell+1)}
G^{I\,\ell-1}_{1\ell, 1\ell}(p, p^\prime; K)
\right.\nonumber \\
& & \left.\ \hspace{10mm}
-\frac{2\ell+3}{(\ell+1)(2\ell+1)}
G^{I\,\ell+1}_{1\ell, 1\ell}(p, p^\prime; K) \right]
\hspace{15mm} (\ell=1,~2,~3,\cdots)\ \ ,
\label{par3}  
\end{eqnarray}
using the transformation formula (see Eq.\,(2.22) of Ref.~\citen{B8a})
\begin{eqnarray}
& &
\ \hspace{-8mm}
u^\Omega_\lambda(k, q; K)=\sum_{i,j}\sum^\infty_{\ell=0
~{\rm or}~1}(2\ell+1)\,F^{\Omega\,\lambda \ell}_{i,j}(k, q)
\,g^\Omega_\ell(p_i, p_j; K)\quad
(\Omega=C,~LS)\ . \hfill
\label{par4}
\end{eqnarray}
Here, $g^\Omega_\ell(p_i, p_j; K)$ is assigned to 
$g^I_{0\,\ell}(p, p^\prime; K)$ or $g^I_{ss\,\ell}(p, p^\prime; K)$
in \eq{par3} for the central part and to $h^I_{0\,\ell}(p, p^\prime; K)$
for the $LS$ part.
In \eq{par3}, the prime on the summation symbol indicates
that the sum is taken only for quantum numbers such as $J$ and $S$
that satisfy the generalized Pauli principle
$(-1)^{\ell+S+I}=-1$, and the angular-momentum coupling $(\ell S)J$.
Similarly, the $LS$ component involves only odd $\ell$ for $I=1$
and even $\ell$ for $I=0$.
As discussed in Ref.\,\citen{B8a}, $u_\lambda(2q^\prime, k/2; K)$ needed
in the calculation of $\CM_{1D_-}(\bk, \bq)
=M_{1D_-}(\bq_f, \bq_i)$ is
obtained by simply changing $g_\ell(p, p^\prime; K)$
to $(-1)^\ell\,g_\ell(p, p^\prime; K)$ in \eq{par4}.
This is a consequence of the symmetry property satisfied
by the coefficients, $F^{\Omega\,\lambda \ell}_{i,j}(k, q)$,
given in Eq.\,(2.23) of Ref.~\citen{B8a}.
Because of this symmetry, we only need to calculate 
the partial wave components for $M_{0D+}(\bq_f, \bq_i)$
for the spin-independent central component $g^I_{0\,\ell}(p, p^\prime; K)$,
as shown in \eq{int13}.

For the $E$-type RGM kernel, only the $S$-wave components of the
$G$-matrix interaction contribute to the RGM kernel:
\begin{eqnarray}
\ \hspace{-8mm} 
\CE(K)= \frac{2}{\pi}\left(\frac{1}{\pi \nu}\right)^{\frac{3}{2}}
\int p^2\,d\,p\,{p^\prime}^2\,d\,p^\prime~\exp \left\{-\frac{1}{2\nu}
\left(p^2+{p^\prime}^2 \right)\right\}
~g_0(p, p^\prime; K)\ ,
\label{par5}
\end{eqnarray}
and
\begin{eqnarray}
E^V_{0E}(q) & = & 4 \pi \left(\frac{1}{2\pi \nu}\right)^{\frac{3}{2}}
\int^\infty_0 K^2\,d\,K\,e^{-\frac{1}{2\nu}\left(K-\frac{1}{2}q\right)^2}
~\widetilde{i}_0\left(\frac{1}{2\nu} q K\right)
\,\CE(K)\ ,\nonumber \\
E^V_{1E}(q) & = & 4 \pi \left(\frac{3}{4\pi \nu}\right)^{\frac{3}{2}}
\int^\infty_0 K^2\,d\,K\,e^{-\frac{3}{4\nu}\left(K-\frac{4}{3}q\right)^2}
~\widetilde{i}_0\left(\frac{2}{\nu}q K\right)
\,\CE(K)\ ,
\label{par6}
\end{eqnarray}
where $\widetilde{i}_\lambda(x)=e^{-x}\,i_\lambda(x)
=e^{-x} i^\lambda j_\lambda(-ix)$ is employed.
For the $0D_+$-type RGM kernel, the partial wave decomposition is
carried out, including the momentum dependence of $K=2|\bq-\bq^\prime|$:
\begin{eqnarray}
\CM_{0D_+}(\bk, \bq) = \sum^\infty_{\lambda=0} 
(2 \lambda+1)~\CM_{0D_+\,\lambda}(k, q)
~P_\lambda(\widehat{\bk}\cdot \widehat{\bq})\ .
\label{par7}
\end{eqnarray}
The result is
\begin{eqnarray}
\CM_{0D_+\,\lambda}(k, q)
& = & 4 \pi~e^{-\frac{3}{32\nu}k^2}
~\left(\frac{8}{3\pi \nu}\right)^{\frac{3}{2}}
\int^\infty_0 {q^\prime}^2\,d\,q^\prime
~\frac{1}{2} \int^1_{-1} d\,x
~e^{-\frac{8}{3\nu}\left(q^\prime-\frac{5}{8}q\right)^2}
\nonumber \\
& & \times 
~e^{-\frac{10}{3\nu}q q^\prime (1-x)}
~u_\lambda \left(k, q^\prime; 2\sqrt{q^2+{q^\prime}^2-2q q^\prime x}\right)
\,P_\lambda(x)\ .
\label{par8}
\end{eqnarray}
This term (and also $M^{0I}_{1E}$ term of \eq{int12}) is 
transformed back to the partial wave component $M_{0D_+\,\ell}(p_f, p_i)$
by the inverse transformation of \eq{par4}.
The partial-wave components of the $LS$ term
$M^{LS}_{0D_+\,\ell}(p_f, p_i)$ are similarly obtained.
For the $S$ type (and also for $S^\prime$ type), it is easiest
to calculate $M_{1S\,\ell}(q_f, q_i)$ directly from
\begin{eqnarray}
M_{1S}(\bq_f, \bq_i) = \sum^\infty_{\ell=0}
(2 \ell+1)~M_{1S\,\ell}(q_f, q_i)
~P_\ell (\widehat{\bq}_f\cdot \widehat{\bq}_i)\ .
\label{par9}
\end{eqnarray}
We obtain
\begin{eqnarray}
\ \hspace{-10mm} & & M_{1S\,\ell}(q_f, q_i) = (-1)^\ell \frac{2}{\pi}
\left(\frac{2}{\nu^2}\right)^{\frac{3}{2}}
~e^{-\frac{1}{16\nu}(3{q_f}^2+7{q_i}^2)}
~\int^\infty_0 {p^\prime}^2\,d\,p^\prime
~e^{-\frac{1}{2\nu}{p^\prime}^2}
\nonumber \\
\ \hspace{-10mm} & & \times \int^\infty_0 K^2\,d\,K
~e^{-\frac{5}{8\nu}K^2+\frac{1}{2\nu}(q_f+q_i) K}
~\widetilde{i}_\ell\left(\frac{1}{2\nu}q_i K\right)
\nonumber \\
\ \hspace{-10mm} & & \times \frac{1}{2}\int^1_{-1} d\,x
~e^{-\frac{1}{2\nu}q_f K (1-x)}
~g_0\left(\sqrt{{q_f}^2+K^2/4-q_f K x}, p^\prime; K\right)
\,P_\ell(x)\ .
\label{par10}
\end{eqnarray}
Here again, we only find the contribution from the $S$-wave components
of the $G$-matrix interaction.
If we modify $\sqrt{{q_f}^2+K^2/4-q_f K x} \rightarrow q_f$
in the third line of \eq{par10}, it becomes
$\widetilde{i}_\ell\left(q_f K/2\nu \right)\,g_0(q_f, p^\prime; K)$.
The $S^\prime$-type component is obtained from the symmetry discussion as
\begin{eqnarray}
M_{1S^\prime\,\ell}(q_f, q_i)=M_{1S\,\ell}(q_i, q_f)\ .
\label{par11}
\end{eqnarray}

With all of these contributions, the partial-wave
component of the RGM Born kernel for the $n \alpha$ system
is given by
\begin{eqnarray}
& & \ \hspace{-5mm} V^{{\rm RGM}\,J}_\ell(q_f, q_i; \varepsilon)
=V^C_\ell(q_f, q_i; \varepsilon)
+V^{LS}_\ell(q_f, q_i)~\langle \bfell \cdot \bS \rangle_{J\ell}
\ ,\nonumber \\ [2mm]
& & \ \hspace{-5mm}
V^C_\ell(q_f, q_i; \varepsilon)
=G^{\rm K}_\ell (q_f, q_i)+\varepsilon K_{\ell}(q_f, q_i)
\nonumber \\
& & \ \hspace{-5mm} \qquad 
+2 \sum^1_{I=0} (2I+1) \left[ M^{0\,I}_{0D+\,\ell}(q_f, q_i)
+M^{0\,I}_{1E\,\ell}(q_f, q_i)
-M^{0\,I}_{1S\,\ell}(q_f, q_i)
-M^{0\,I}_{1S\,\ell}(q_i, q_f) \right]\ ,\nonumber \\
& & \ \hspace{-5mm} V^{LS}_\ell(q_f, q_i)
=4 \sum^1_{I=0} (2I+1) M^{LS\,I}_{0D+\,\ell}(q_f, q_i)\ ,
\label{par12}
\end{eqnarray}
where $\langle \bfell \cdot \bS \rangle_{J\ell}=\ell/2$ for
$J=\ell+1/2$ and $-(\ell+1)/2$ for $J=\ell-1/2$, and
$K_\ell(q_f, q_i)$ and $G^{\rm K}_\ell (q_f, q_i)$ are given
in \eq{a3}.

\subsection{Selection of starting energies of $G$-matrix}

In previous calculations of $B_8 \alpha$ interactions
with $B_8=\Lambda,~\Sigma$ and $\Xi$, we used the starting-energy
dependence of the $G$-matrix according to the rule of the 
nuclear matter calculation. Namely, we used the angle-averaging
procedure over the relative momentum $\bq$ (which is the integral variable)
in the relationship, $\bK=\bq_1+\bq_2$ and $\bq=(1/2)(\bq_1-\bq_2)$,
under the constraint $|\bq_2|<k_F$. This procedure yields
$K=2[{q_1}^2+q^2-q_1 q (1+[-1|x|1])]^{1/2}$ and
$q_2=\sqrt{K^2/2+2 q^2-{q_1}^2}$. Here, $x$ is defined by
$x=({q_1}^2+4q^2-{k_F}^2)/(4q_1 q)$
and $[-1|x|1]={\rm max}\{-1, {\rm min}\{x,1\}\}$.
The starting-energy $\omega$ is then calculated from
\begin{eqnarray}
\omega=\frac{\hbar^2}{M_N}q^2+U_N(q_1)+U_N(q_2)\ ,
\label{stat1}
\end{eqnarray}
after the subtraction of the conserved c.m.~energy $(\hbar^2/2M_N)K^2$.
The assignments $\bq$ to $\bq_1$ and ${\bq}^\prime$ to $\bq$ in \eq{int15}
(and similar prescription in the $1S$- and $1S^\prime$ terms
in \eq{int16}) clearly destroy the redundancy property
of the $n \alpha$ RGM kernel.
Although the $K$ dependence is explicitly treated in the partial wave
decomposition, we need at least one more vector to uniquely specify
the momenta $q_1$ and $q_2$ even under the assumption
of the angle-averaging procedure.
Here we use a set of $\bK$ and $\bq$ to specify the starting energy $\omega$.
This method, however, involves some ambiguity,
since the definitions of $\bq_1$ and $\bq$ differ for each type.
The explicit expressions in Eqs.\,(\ref{int15})-(\ref{int16}) indicate
that the roles of $\bq$ and $\bq^\prime$ in the $0D_+$ and $1D_-$ types
are taken over by $\bq_f$ and $\bp$ in the $1S$ type and
by $\bq_i$ and $\bp^\prime$ in the $1S^\prime$ type.
Such a choice of three different sets of the $G$-matrix interaction
parameters apparently causes a problem. Here, we choose
the $0D_+$-type definitions for $\bq_1$ and $\bq$ as a standard set.
Namely, we assume that \eq{int5} is explicitly given by
\begin{eqnarray}
\ \hspace{-8mm}
\langle \bp_1, \bp_2 | u | \bp^\prime_1, \bp^\prime_2 \rangle
=\delta(\bK-\bK^\prime)\frac{1}{(2\pi)^3}
g\left(\bp, \bp^\prime;
\omega\left(\vert(\bp+\bp^\prime)/2\vert, K\right), K, k_F\right)\ .
\label{stat2}
\end{eqnarray}
In practical calculation, we first assume $q=|\bq|$ and
$K=|\bK|$ and calculate the $G$-matrix using the angle-averaging
procedure of $\bK$ under the constraint $|\bK/2-\bq|< k_F$.
Then the rule of the partial-wave decompositions in $g$ and $u$ is
given by
\begin{eqnarray}
& & g\left(\bp, \bp^\prime;
\omega\left(q, K\right), K, k_F\right)
=\sum_\ell (2\ell+1)\,g_\ell(p, p^\prime; \omega(q,K), K, k_F)
\,P_\ell(\widehat{\bp}, \widehat{\bp}^\prime)
\nonumber \\
& & =u(\bk^\prime, \bq^\prime; \omega(q, K), K, k_F)
\nonumber \\
& & =\sum_\lambda (2 \lambda+1)\,u_\lambda(k^\prime, q^\prime;
\omega(q, K), K, k_F)\,P_\lambda(\widehat{\bk}^\prime,
\widehat{\bq}^\prime)\ .
\label{stat3}
\end{eqnarray}
We pick up the $q=q^\prime$ portion in the last line of \eq{stat3}
and define
\begin{eqnarray}
\widetilde{u}_\lambda(k^\prime, q^\prime; K, k_F)
=u_\lambda(k^\prime, q^\prime; \omega(q^\prime, K), K, k_F)\ .
\label{stat4}
\end{eqnarray}
The full $G$-matrix interaction is constructed using
\begin{eqnarray}
& & \widetilde{u}(\bk^\prime, \bq^\prime; K, k_F)
=\sum_\lambda (2\lambda+1)\,\widetilde{u}_\lambda(k^\prime, q^\prime;
K, k_F)\,P_\lambda(\widehat{\bk}^\prime,
\widehat{\bq}^\prime)\nonumber \\
& & =\widetilde{g}(\bp, \bp^\prime; K, k_F)
=\sum_\ell (2\ell+1)\,\widetilde{g}_\ell(p, p^\prime; K, k_F)
\,P_\ell(\widehat{\bp}, \widehat{\bp}^\prime)\ ,
\label{stat5}
\end{eqnarray}
which is equivalent to
\begin{eqnarray}
g\left(\bp, \bp^\prime;
\omega\left(q^\prime, K\right), K, k_F\right)
=u(\bk^\prime, \bq^\prime; \omega(q^\prime, K), K, k_F)\ .
\label{stat6}
\end{eqnarray}
Since the $q^\prime$ dependence in the starting-energy part
is absorbed in the relative momenta in the bra and ket sides
of the $G$-matrix, we can apply the previous whole formalism to the
$G$-matrix interactions $\widetilde{u}$ and $\widetilde{g}$.
Note that we only need the $S$-wave component for $\widetilde{g}$,
since only this component is required
for the $0E$, $1E$, $1S$ and $1S^\prime$ types.    

The angle average over $\bK$ in the $G$-matrix calculation
is carried out similarly to that in the previous case. We start from
\begin{eqnarray}
\omega(\bq, \bK)=\frac{\hbar^2}{M_N}q^2
+U_N\left(\bK/2+\bq \right)
+U_N\left(\bK/2-\bq \right)\ ,
\label{stat7}
\end{eqnarray}
and make the angle-averaging over $\bK$ under the constraint
$|\bK/2-\bq|< k_F$. Then we find
\begin{eqnarray}
\omega(q, K) & = & \frac{\hbar^2}{M_N}\,q^2
+U_N\left(\sqrt{K^2/4+q^2+K q \langle x \rangle} \right)
\nonumber \\
& & +U_N\left(\sqrt{K^2/4+q^2-K q \langle x \rangle} \right)\ ,
\label{stat8}
\end{eqnarray}
where
\begin{eqnarray}
\langle x \rangle
=\frac{1}{2}(1+[-1|x|1])
=\left\{ \begin{array}{c}
1 \\
\frac{1}{2}(1+x) \\
0 \\
\end{array}\right.
\quad \hbox{for} \quad
\left\{ \begin{array}{c}
x >1 \\
-1 < x < 1 \\
x < -1 \\
\end{array} \right. \ ,
\label{stat9}
\end{eqnarray}
with
\begin{eqnarray}
x = \frac{1}{Kq}\left(\frac{1}{4}K^2+q^2-{k_F}^2\right)\ .
\label{stat10}
\end{eqnarray}
In two special cases, we find
\begin{eqnarray}
\omega (q, K)=\left\{ \begin{array}{c}
\frac{\hbar^2}{M_N}q^2
+U_N(\sqrt{K^2/4+q^2})+U_N(\sqrt{K^2/4+q^2}) \\ [2mm]
\frac{\hbar^2}{M_N} q^2
+U_N(K/2+q)+U_N(|K/2-q|) \\ [2mm]
\end{array} \right.
\nonumber \\
\hspace{50mm} \hbox{for} \quad
\left\{ \begin{array}{c}
K/2+q < k_F \\ [2mm]
\vert K/2-q \vert >k_F \\
\end{array} \right.\ .
\label{stat11}
\end{eqnarray}

\section{Results and discussion}

\subsection{$NN$ $G$-matrix and $n\alpha$ phase shifts}

\begin{figure}[htb]
\begin{center}
\includegraphics[width=0.49\textwidth]{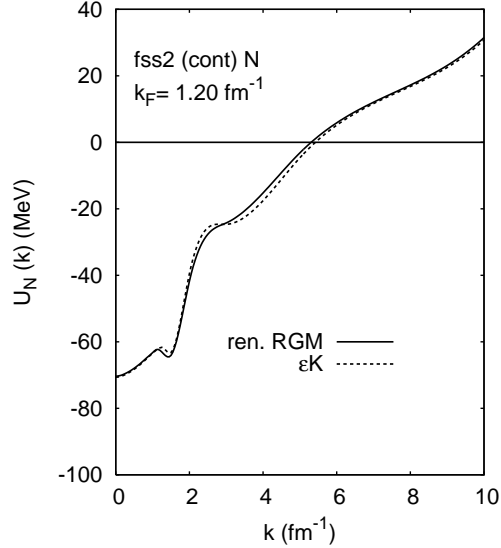}
\caption{
Comparison of the nucleon s.p.~potentials predicted by the
renormalized RGM kernel and the $\varepsilon K$ prescription. The model
is fss2 and $k_F=1.20~\hbox{fm}^{-1}$ is assumed in the
continuous prescription for intermediate spectra.
}
\label{fig1}
\end{center}
\end{figure}

Figure \ref{fig1} shows the nucleon s.p.~potential of the model fss2,
obtained by $G$-matrix calculation for symmetric nuclear matter.
The Fermi momentum $k_F=1.20~\hbox{fm}^{-1}$ is assumed in
the continuous prescription for intermediate spectra.
The solid curve (ren.~RGM) is obtained from the renormalized RGM kernel
and the dashed curve ($\varepsilon K$) from the energy-dependent
RGM kernel with the explicit $\varepsilon K$ term.
We find that the off-shell transformation by $1/\sqrt{N}$ gives
a rather minor modification for the $NN$ $G$-matrix, giving a slightly
repulsive effect to the nucleon s.p.~potential. We use the prescription
of the renormalized RGM to calculate
the $n \alpha$ Born kernel in this paper.
The internal energy of the $\alpha$ cluster, calculated from \eq{int14}
with $q=0$ and $\nu=0.257~\hbox{fm}^{-2}$, is $E_\alpha(0)=-26.5$ MeV.

\begin{figure}[htb]
\begin{center}
\begin{minipage}[h]{0.45\textwidth}
\includegraphics[width=\textwidth]{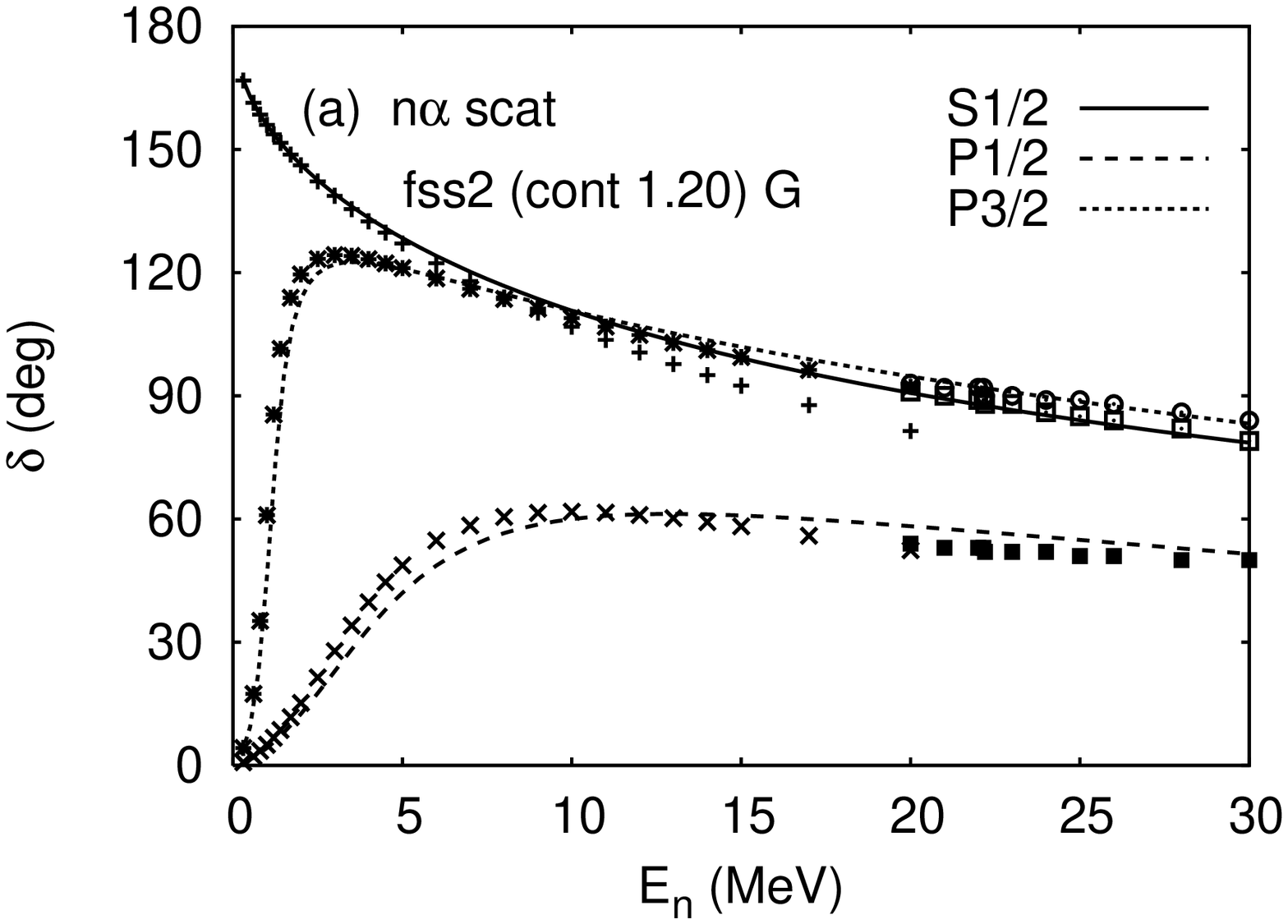}
\includegraphics[width=\textwidth]{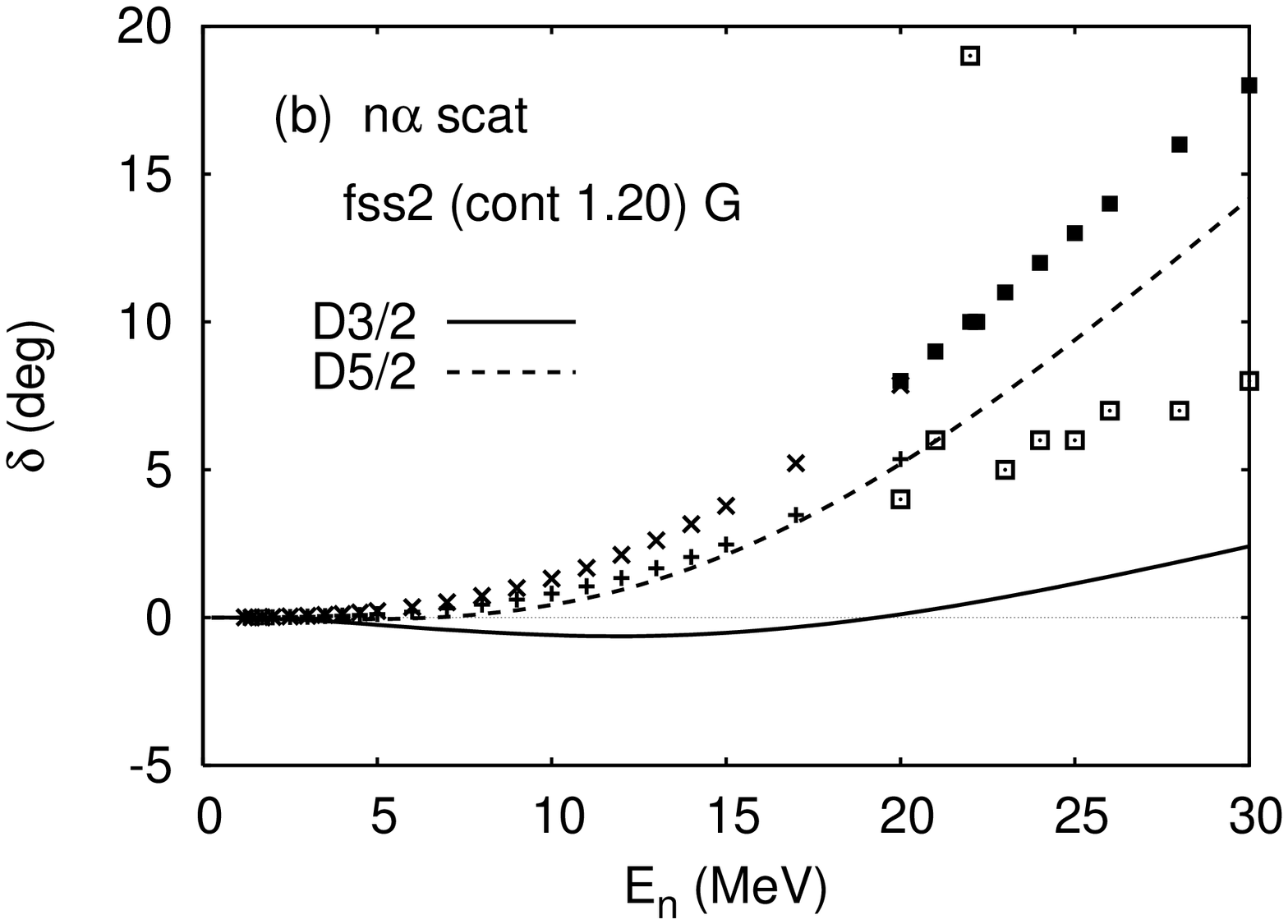}
\includegraphics[width=\textwidth]{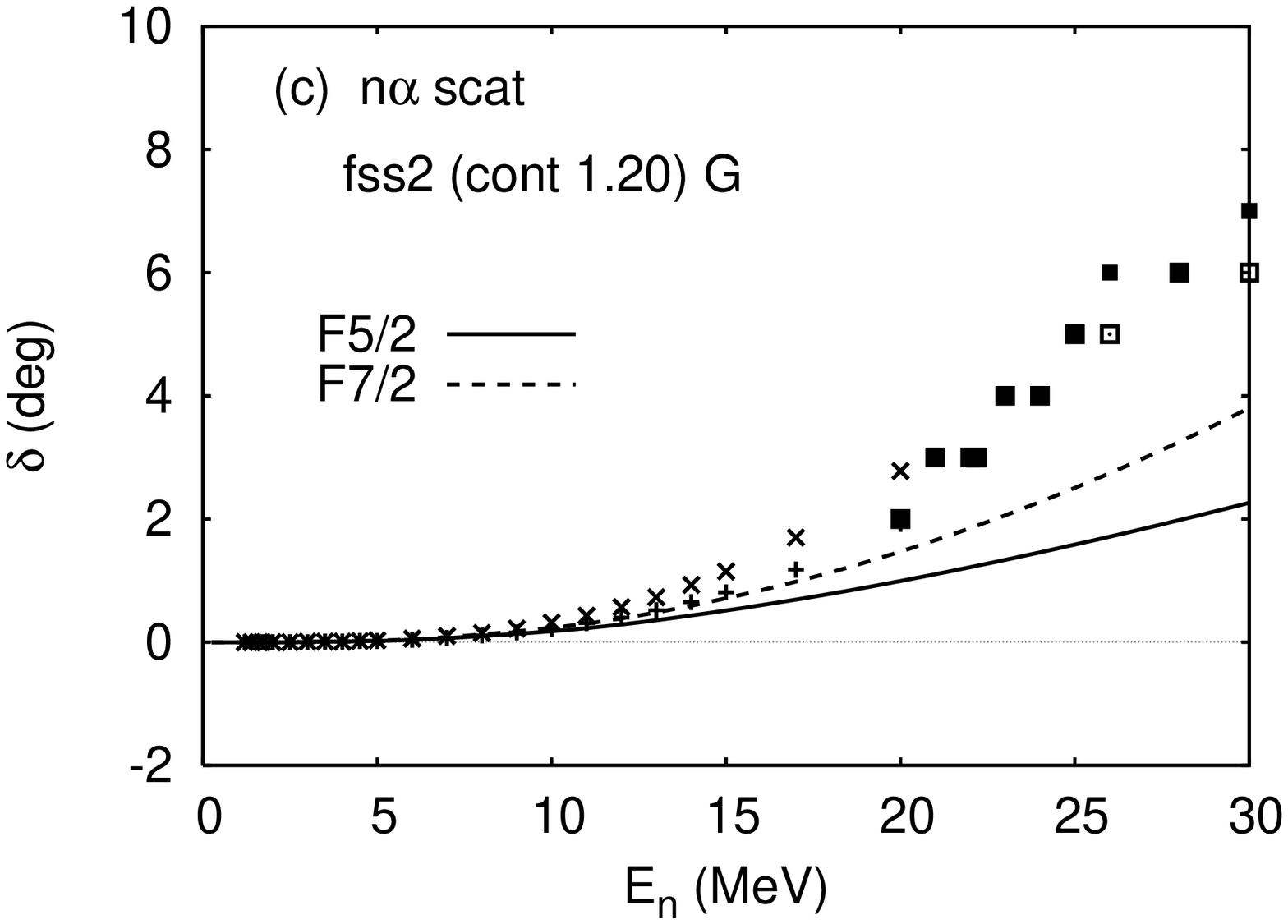}
\caption{
$n \alpha$ RGM phase shifts predicted by the quark-model
$G$-matrix interaction by fss2
for $k_F=1.20~\hbox{fm}^{-1}$.
(a) $S_{1/2}$, $P_{3/2}$ and $P_{1/2}$ states.
(b) $D_{5/2}$ and $D_{3/2}$ states.
(c) $F_{7/2}$ and $F_{5/2}$ states.
The $(0s)^4$ shell-model wave function with the h.o.~size
parameter $\nu=0.257~\hbox{fm}^{-2}$ is used
for the $\alpha$-cluster wave function.
The experimental data are taken from Refs.~\citen{SW72} and \citen{HB66}.}
\label{fig2}
\end{minipage}
\hfill
\begin{minipage}[h]{0.45\textwidth}
\vspace{-19.5mm}
\includegraphics[width=\textwidth]{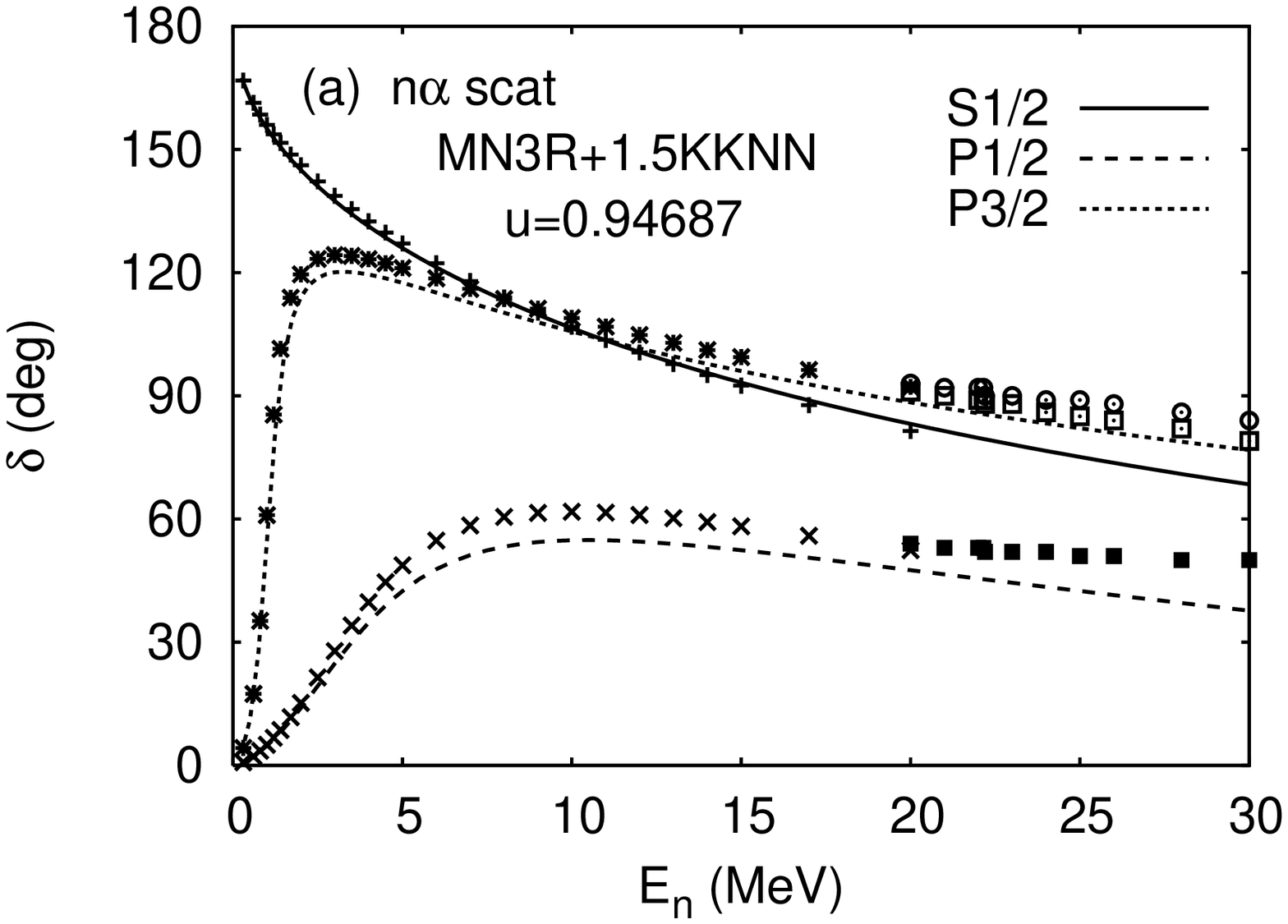}
\includegraphics[width=\textwidth]{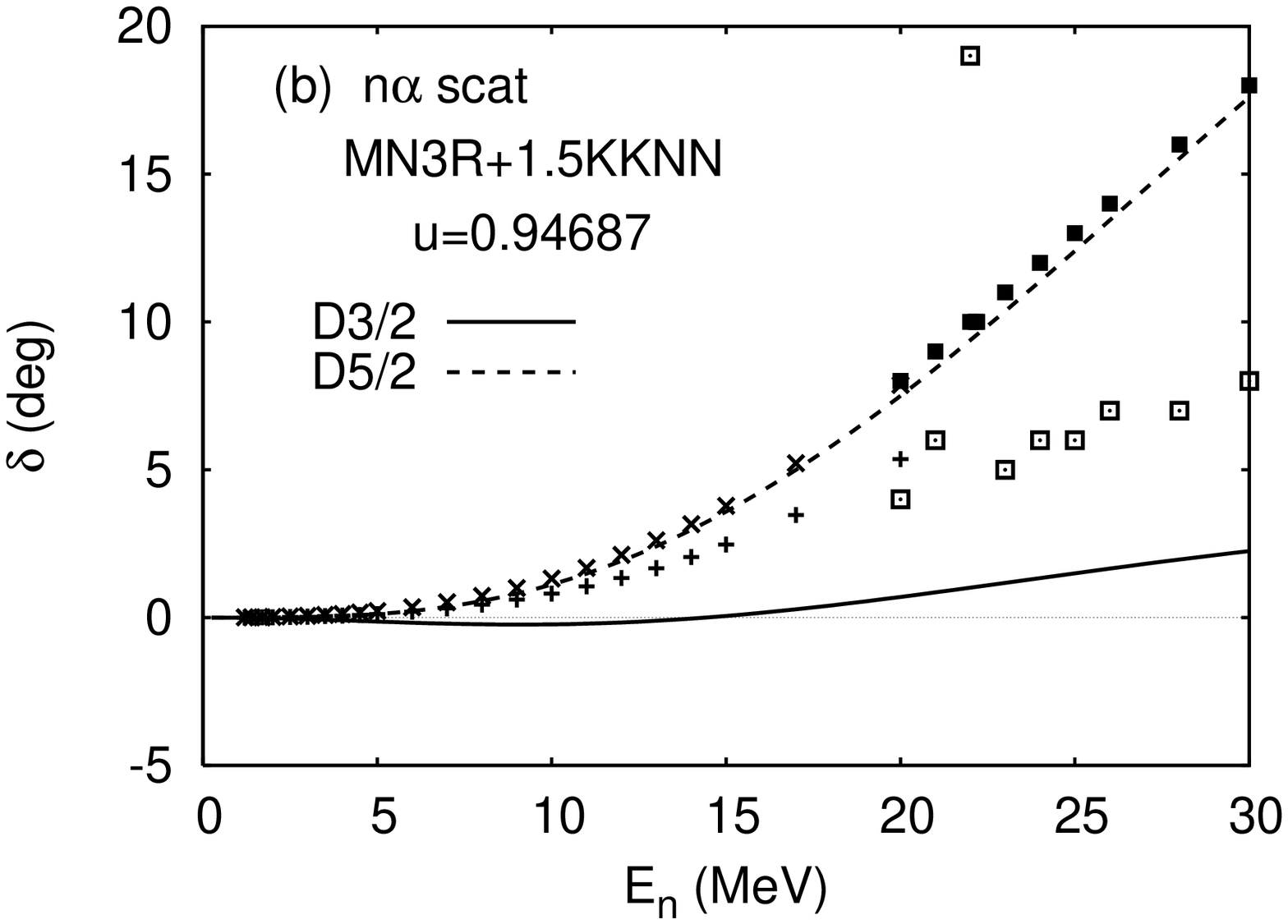}
\includegraphics[width=\textwidth]{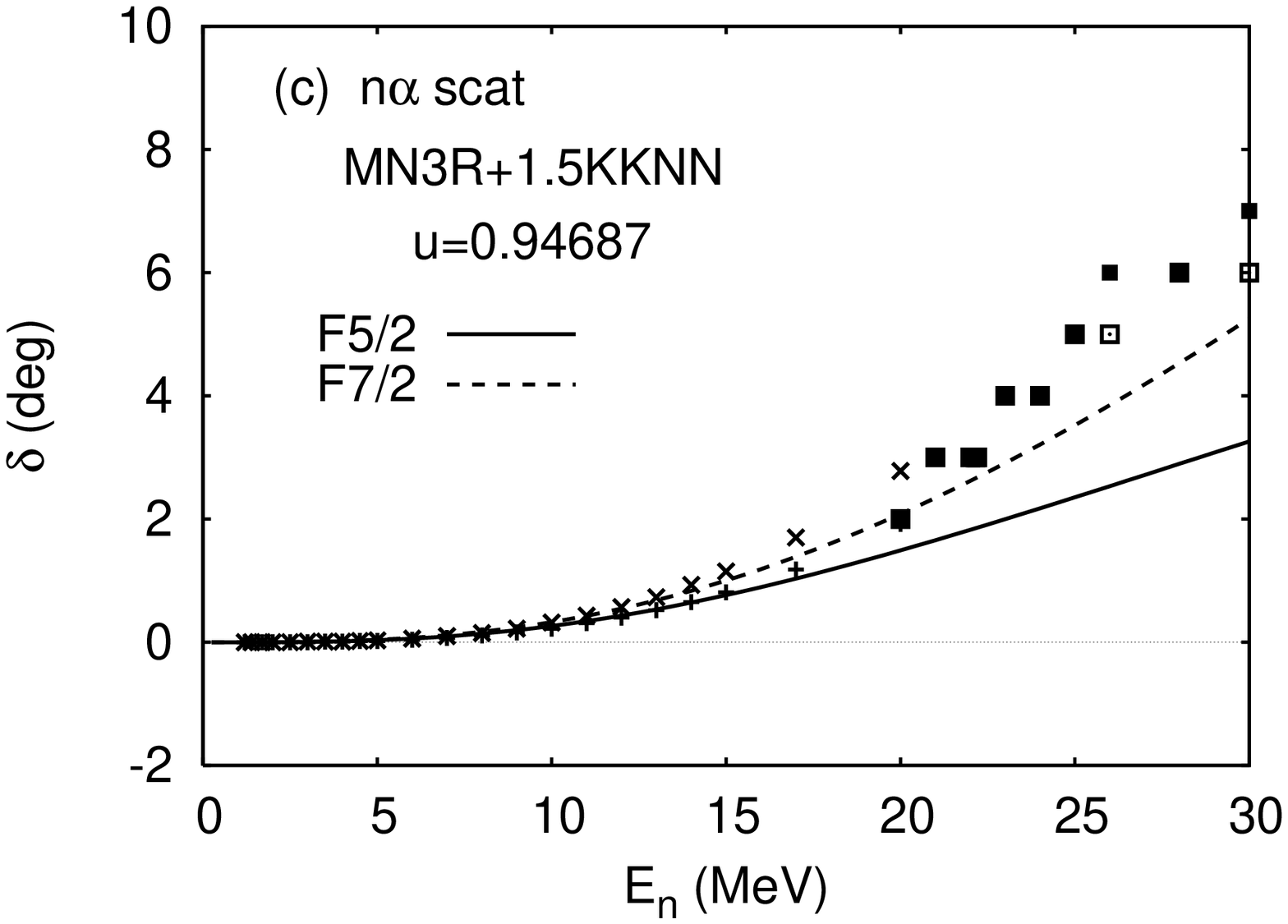}
\caption{
$n \alpha$ RGM phase shifts predicted by the Minnesota 3-range force
with $u=0.94687$ and $\nu=0.257~\hbox{fm}^{-2}$.
The $LS$ force is KKNN $\times 1.5$. (See the text.)
}
\label{fig3}
\end{minipage}
\end{center}
\end{figure}

Figure \ref{fig2} shows the $n \alpha$ phase shifts
of some low partial waves,
predicted by fss2, for the neutron incident energies,
$E_n=0$ - 30 MeV. The $G$-matrix calculation
in symmetric nuclear matter is carried out
using the s.p.~potential in Fig.~\ref{fig1} and
the prescription for the starting energies discussed
in Sec.~2.4. The Fermi momentum $k_F=1.20~\hbox{fm}^{-1}$ corresponds
to 70$\%$ of the normal saturation density ($0.7\,\rho_0$).
The h.o.~width parameter $\nu=0.257~\hbox{fm}^{-2}$
used for the $(0s)^4$ $\alpha$ cluster reproduces the
root-mean-square (rms) radius of the $\alpha$ particle.
Phase-shift solutions are obtained by solving the Lippmann-Schwinger
equation for the $n \alpha$ Born kernel in \eq{par12}.
In Fig.~\ref{fig2}(a), we find that the $n \alpha$ phase shifts
in $L_J=S_{1/2}$, $P_{3/2}$ and $P_{1/2}$ states are well
reproduced, although the $S_{1/2}$ phase shift might be slightly 
too attractive at higher energies. For higher partial waves
with $D_{5/2, 3/2}$ and $F_{7/2, 5/2}$ in Figs.~\ref{fig2}(b) and (c),
the central attraction is not sufficiently large,
although the spin-orbit splitting seems to be reasonably good.
The $D_{3/2}$ phase shift is particularly too low in Fig.~\ref{fig2}(b)
(and also in Fig.~\ref{fig3}(b)).
Part of the reason for the missing attraction
in the $D_{3/2}$ channel is because the specific distortion effect
of the $d+\hbox{}^3\hbox{H}$ channel is not included in the
calculation, whose threshold opens at $E_n \sim -22$ MeV.

For comparison, we show in Figs.~\ref{fig3}(a) - (c) the $n\alpha$ phase
shifts obtained by the standard RGM calculation, using an
effective $NN$ force. In this calculation, the Minnesota three-range
force (MN3R) \cite{TH77} with the Majorana exchange mixture $u=0.94687$
is used for the central force.
The $\alpha \alpha$ phase shifts are well reproduced
in the $\alpha \alpha$ RGM calculation
using this effective $NN$ force \cite{2al}.
For the $LS$ force, the two-range Gaussian $LS$ force
by Kanada {\rm et al.} \cite{KA79},
is used with the spin-isospin coefficients $W=0.5$ and $H=-0.5$
(no $\hbox{}^3E$ $LS$).
If we use the force parameters given in Table I of Ref.~\citen{KA79},
the $LS$ splitting of the $P_{3/2}$ and $P_{1/2}$ states becomes too small.
This is because these authors introduced the $D$-wave component
of the $\alpha$ cluster and an extra contribution to the $LS$ splitting
originates from the two-nucleon tensor force.
If we multiply the strength of this $LS$ force by a factor 1.5,
it gives the correct magnitude, as seen in Fig.~\ref{fig3}(a).
We call this set of effective $NN$ forces MN3R+1.5KKNN.
Although the results in Figs.~\ref{fig2} are not as good
as those of MN3R+1.5KKNN in Fig.~\ref{fig3},
it is clear that our quark-model $NN$ interaction gives a reasonable
description of the $n \alpha$ scattering through the
$G$-matrix approach of the bare interaction.

\bigskip

\subsection{Scattering cross sections and polarization}

\begin{figure}[b]
\begin{center}
\includegraphics[width=0.45\textwidth]{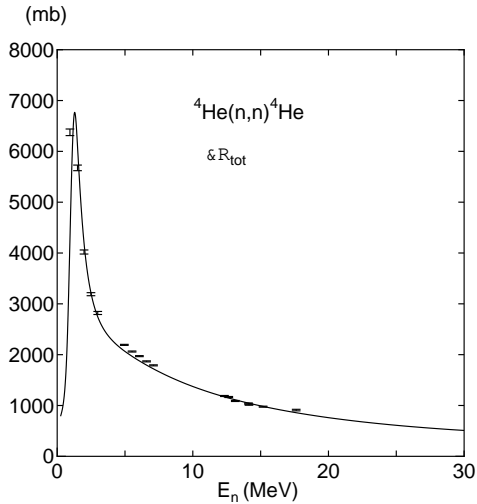}
\caption{
Calculated $n \alpha$ total cross sections by fss2,
compared with the experiment.
The experimental data are taken from Ref.~\citen{Ba59}. 
}
\label{fig4}
\end{center}
\end{figure}

\begin{figure}[htb]
\begin{center}
\begin{minipage}[h]{0.38\textwidth}
\includegraphics[width=\textwidth]{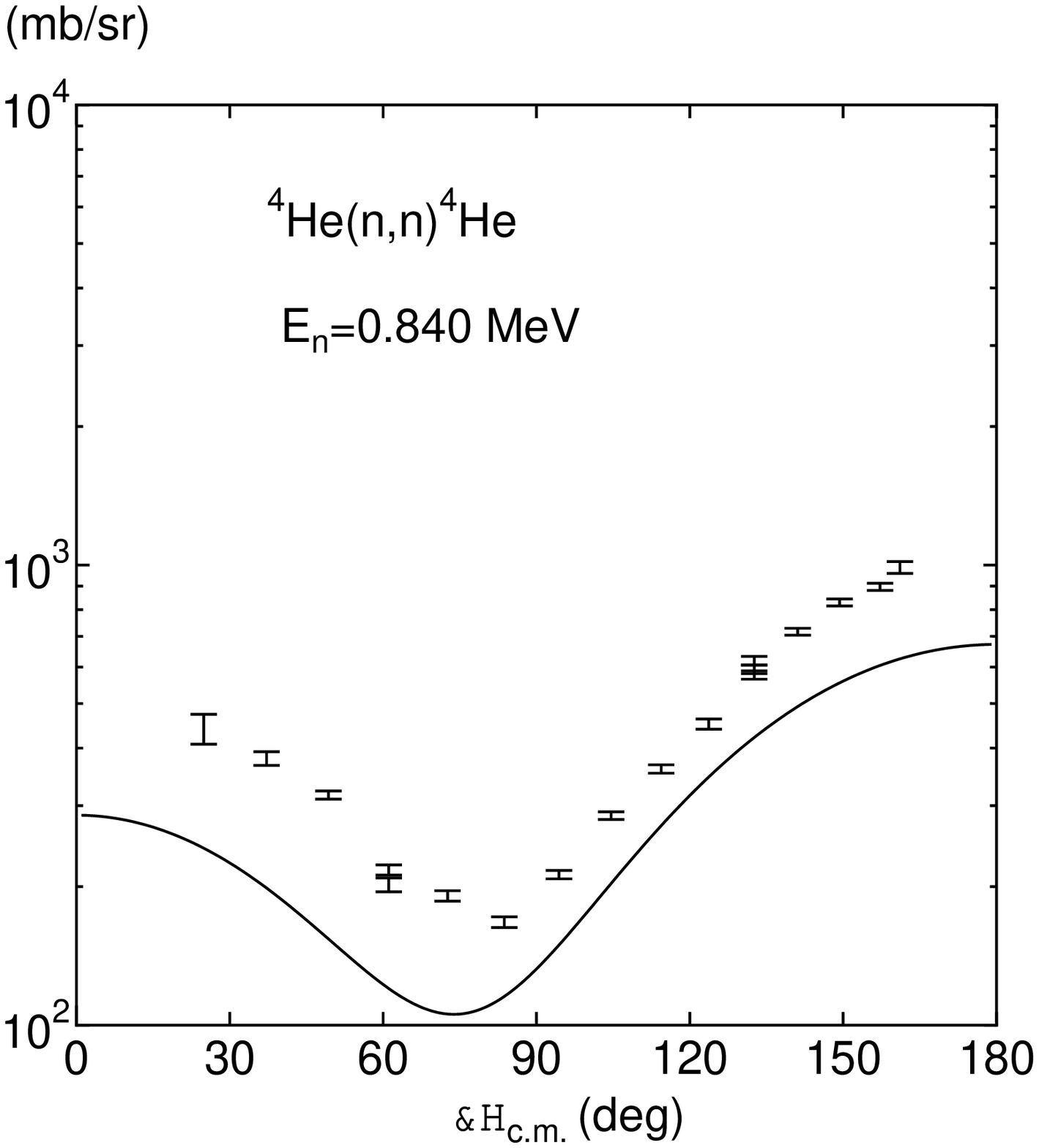}
\includegraphics[width=\textwidth]{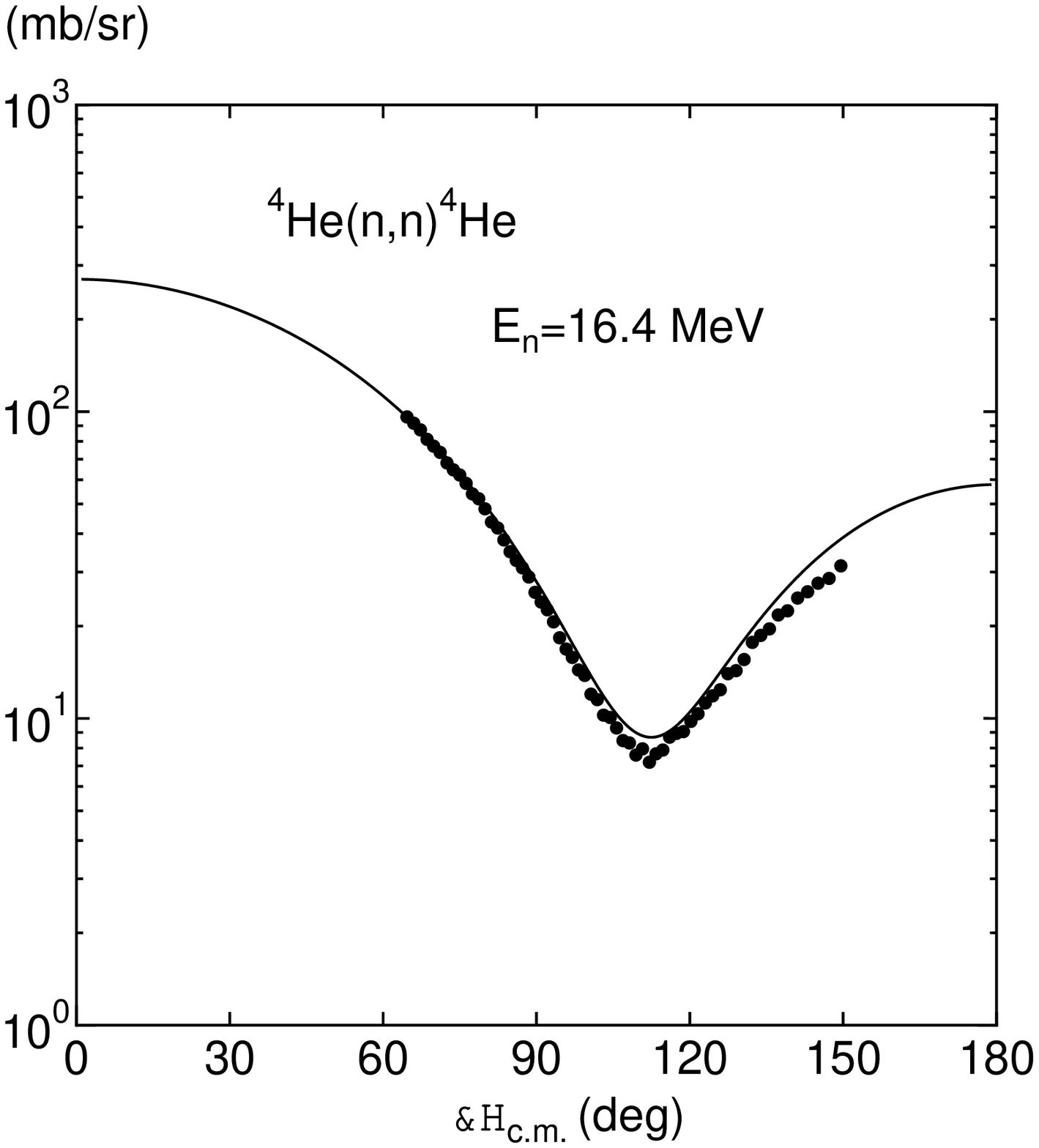}
\includegraphics[width=\textwidth]{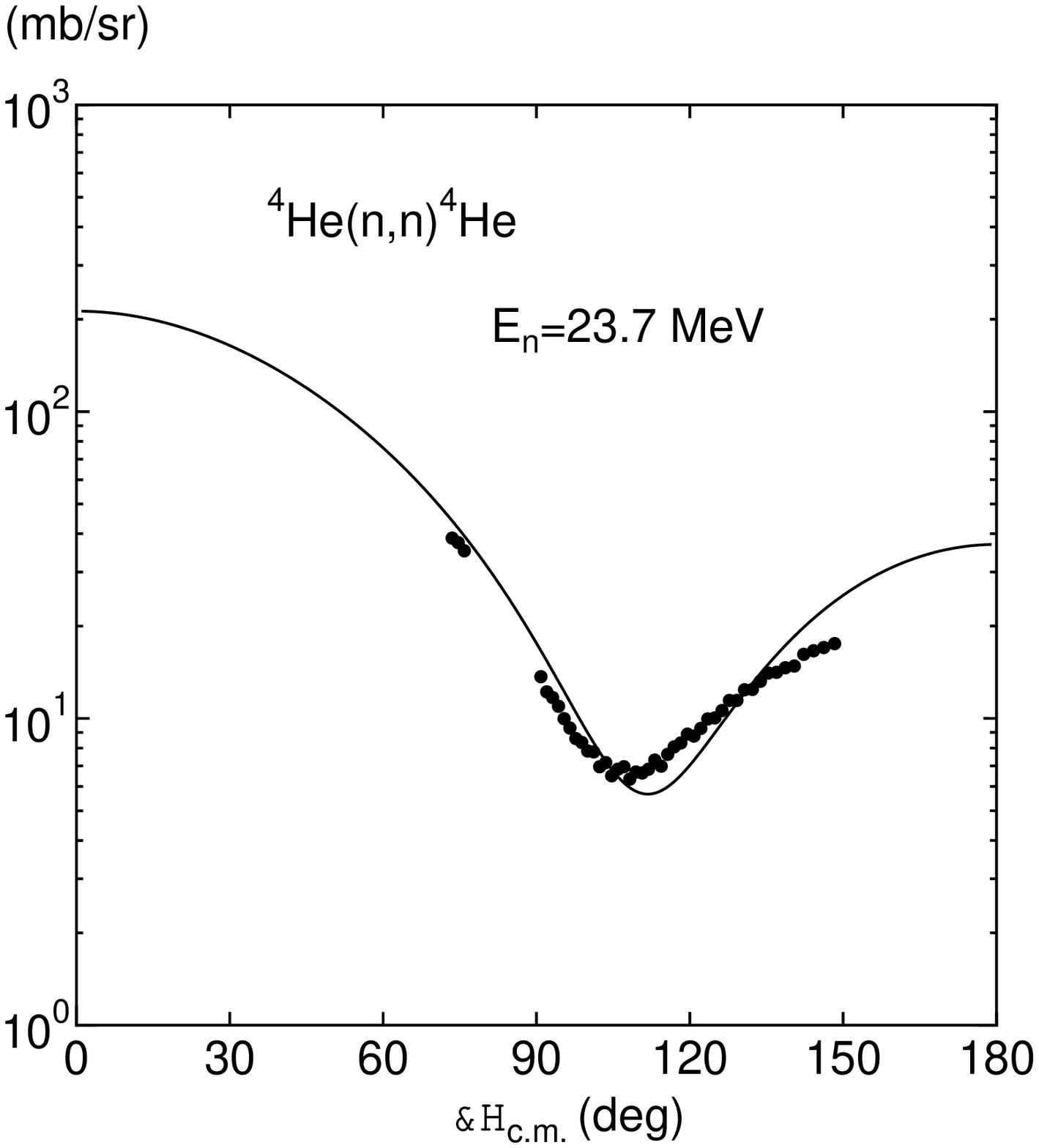}
\caption{
Differential cross sections of the $n \alpha$ scattering
predicted by fss2.
The experimental data are taken from Refs.~\citen{Cr72} and \citen{Sh63}.
}
\label{fig5}
\end{minipage}
\hfill
\begin{minipage}[h]{0.4\textwidth}
\includegraphics[width=\textwidth]{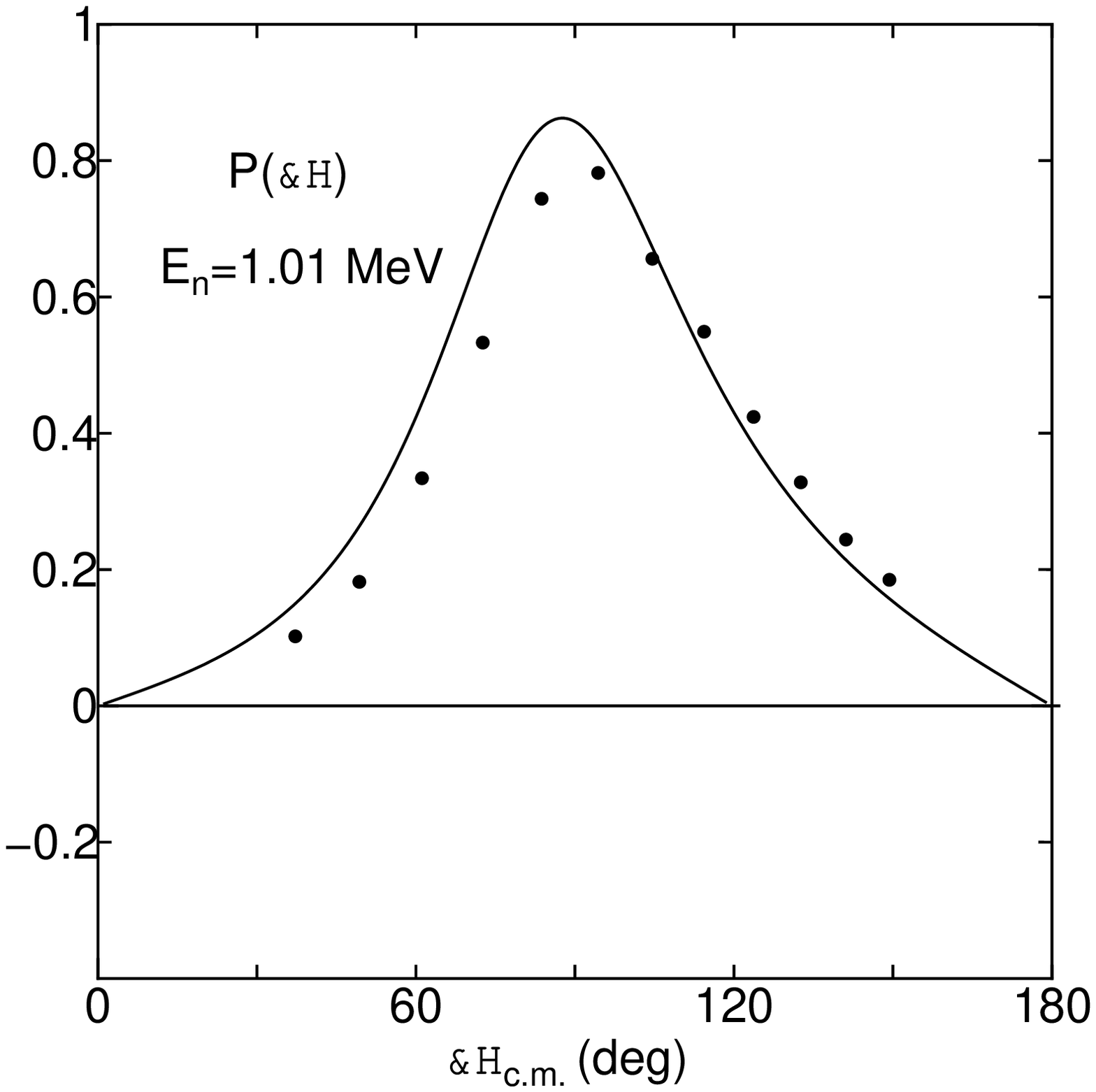}
\includegraphics[width=\textwidth]{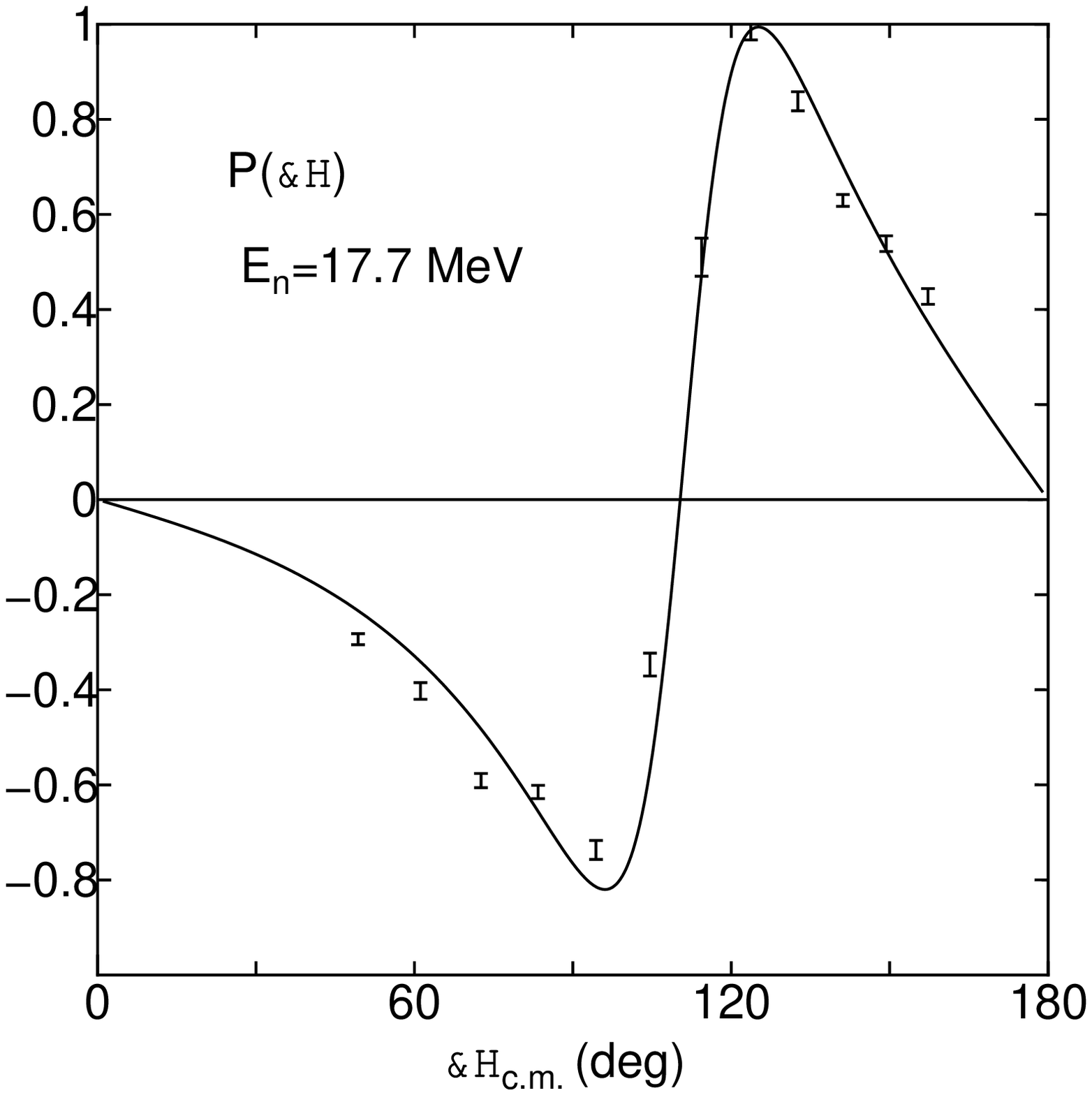}
\includegraphics[width=\textwidth]{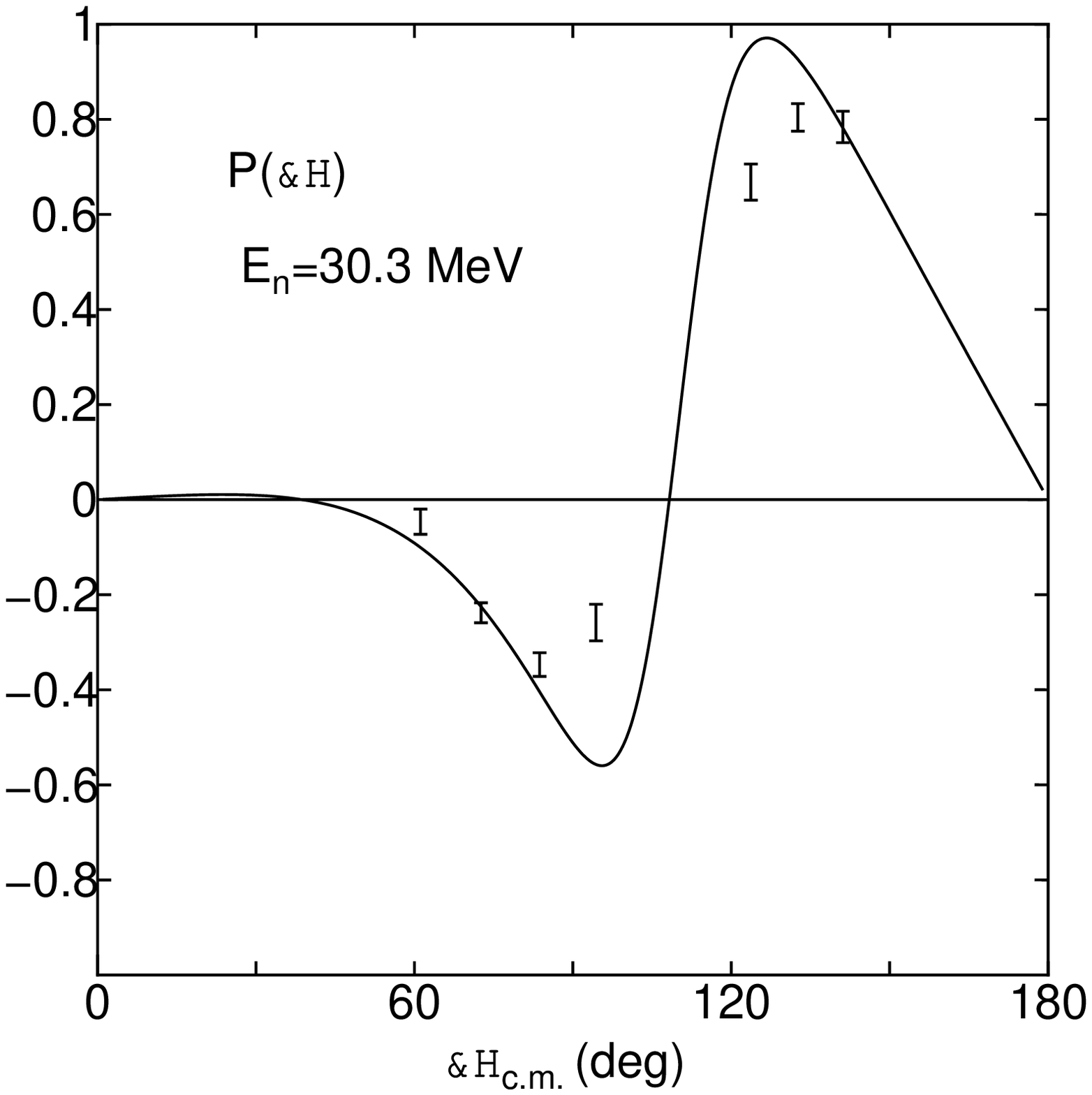}
\caption{
Polarization of the $n \alpha$ scattering predicted by fss2.
The experimental data are taken from Refs.~\citen{Sa68} and \citen{Br72}.
}
\label{fig6}
\end{minipage}
\end{center}
\end{figure}

Since the agreement of the calculated phase shifts with the empirical
values determined from the $R$-matrix analyses \cite{SW72,HB66} is
not complete, we examine the scattering cross sections and polarization
of the $n \alpha$ scattering directly with the experiment.
Figure \ref{fig4} shows the $n \alpha$ total cross sections
up to $E_n=30$ MeV. The prominent peak structure at $E_n=1$ - 2 MeV
is due to the sharp $P_{3/2}$ resonance. Although the calculated
results at around $E_n=4$ - 6 MeV are slightly small, the agreement
from 12 to 18 MeV is satisfactory. We also examine the differential
cross sections and polarization in Figs.~\ref{fig5} and \ref{fig6}
at some available energies. We obtain a fare agreement between
the calculational and experimental results
except in the threshold energy region,
$E_n \sim 22$ MeV, for the decay to the $d+\hbox{}^3\hbox{H}$ channel.
For higher energies, we need to introduce imaginary potentials.

\bigskip

\subsection{Analysis of Born amplitudes}
\bigskip

\begin{figure}[b]
\begin{center}
\begin{minipage}[h]{0.45\textwidth}
\includegraphics[width=\textwidth]{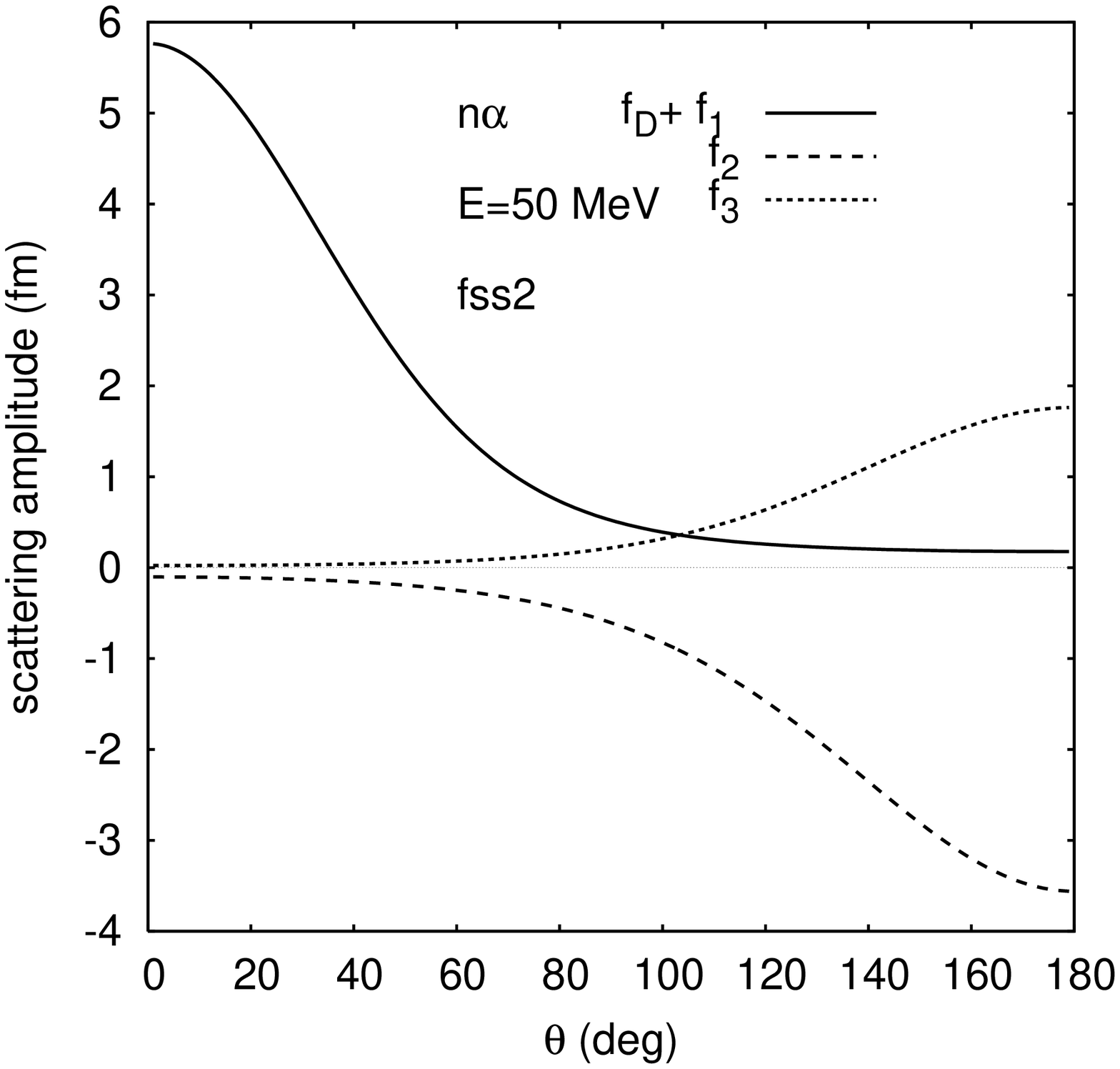}
\includegraphics[width=\textwidth]{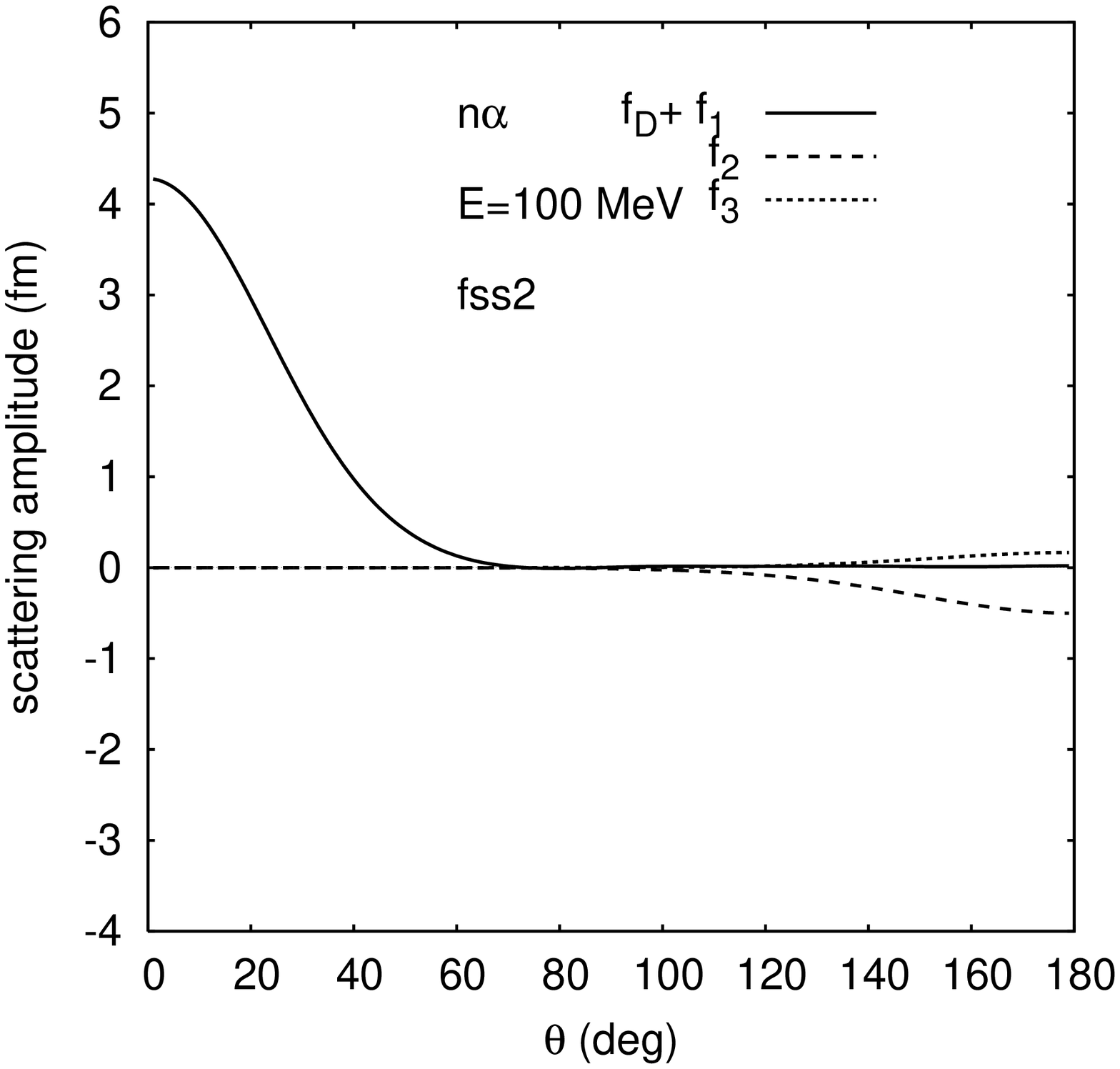}
\caption{
$n \alpha$ Born amplitudes of the $G$-matrix $NN$ interaction
by fss2 with $k_F=1.20~\hbox{fm}^{-1}$ and $\nu=0.257~\hbox{fm}^{-2}$
at the c.m.~energies $E_{\rm c.m.}=50$ and 100 MeV.
}
\label{fig7}
\end{minipage}
\hfill
\begin{minipage}[h]{0.45\textwidth}
\includegraphics[width=\textwidth]{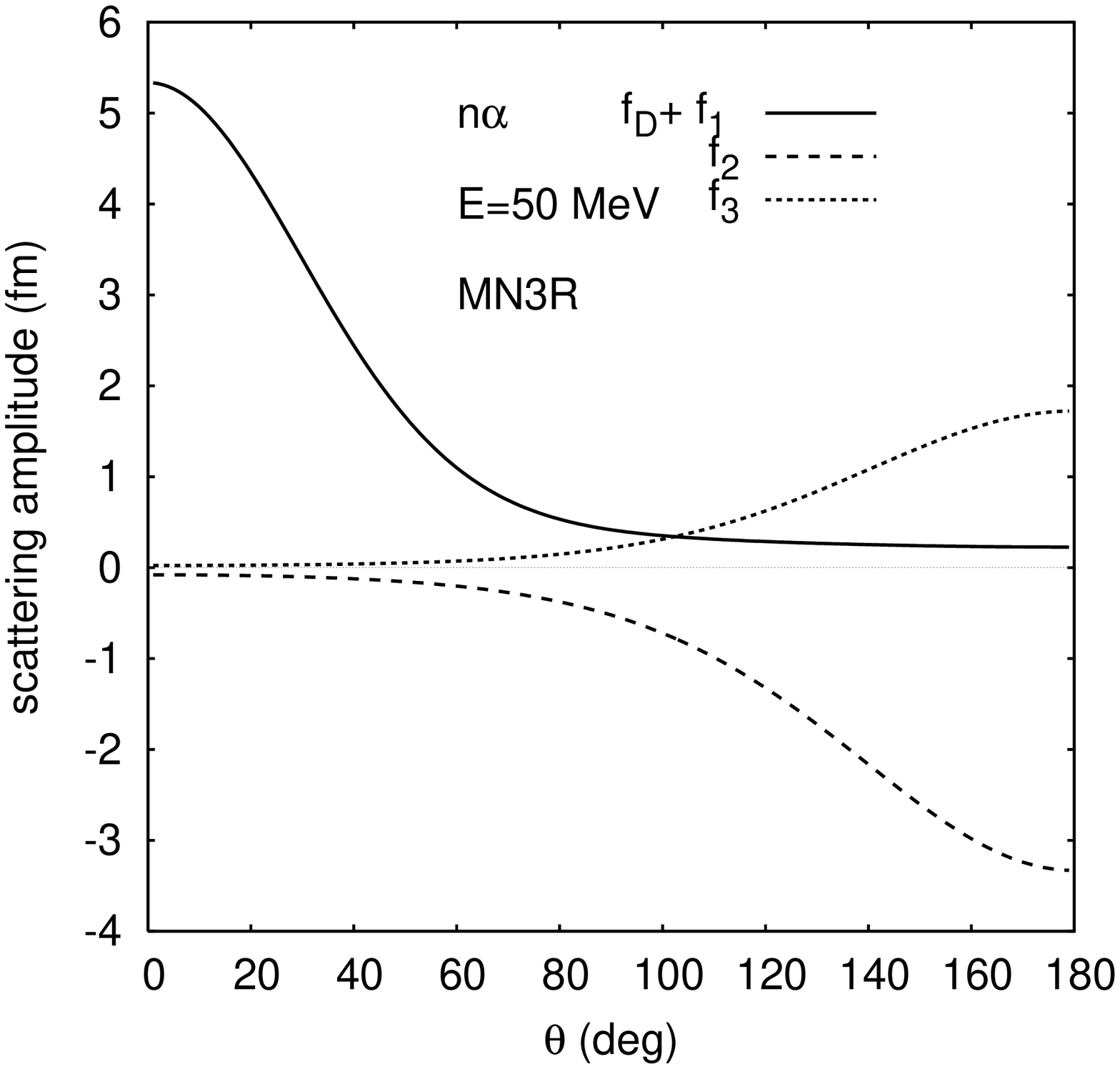}
\includegraphics[width=\textwidth]{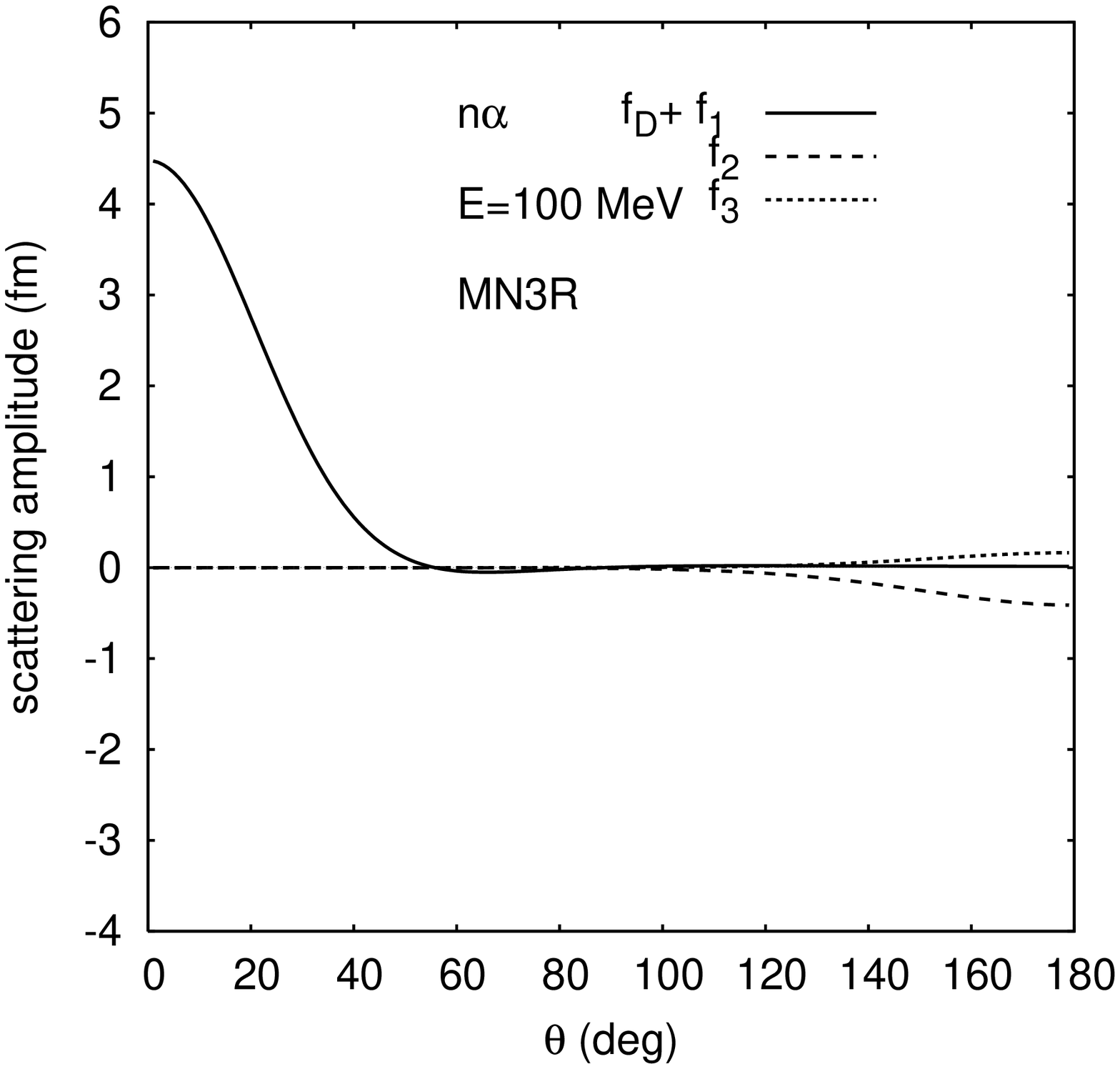}
\caption{
$n \alpha$ Born amplitudes of the MN3R force
with $u=0.94687$ and $\nu=0.257~\hbox{fm}^{-2}$ at the 
c.m.~energies $E_{\rm c.m.}=50$ and 100 MeV.
}
\label{fig8}
\end{minipage}
\end{center}
\end{figure}


Since we calculated the $n \alpha$ Born kernel using only
the real part of the $G$-matrix, the success of the $n \alpha$
single-channel RGM calculation is qualitative
for energies greater than 30 MeV.
Here, we analyze the roles of the various exchange terms of the RGM
central kernels from the $G$-matrix $NN$ interaction.
Such an analysis was carried out by Thompson and Tang \cite{TT71}
for a simple one-term Gaussian $NN$ interaction.
They found that three different groups of the nuclear exchange 
terms have characteristic behavior of various reaction processes.
First, the direct ($0D_+$-type) and knock-on ($1D_-$-type) terms,
specified by $f_D$ and $f_1$ terms in the Born amplitudes below,
respectively, have large amplitudes at the forward angle,
$\theta=0$, characteristic of the Wigner-type potential.
Secondly, the heavy-particle pickup ($1S$ and $1S^\prime$) term
$f_2$ and the nucleon-rearrangement amplitude $f_3$ are important
at backward angles. They are mainly responsible for the existence  
of large scattering cross sections at these angles. In the $G$-matrix
formalism, these Born amplitudes are given by
\begin{eqnarray}
& & f_D+f_1=-\frac{\mu}{2\pi \hbar^2}\,2\,\sum^\infty_{\ell=0}
\sum^1_{I=0} (2\ell+1) (2I+1)\,M^{0\,I}_{0D+\,\ell}(q, q)
\,P_\ell(\cos\,\theta)\ ,\nonumber \\
& & f_2=\frac{\mu}{2\pi \hbar^2}\,4\,\sum^\infty_{\ell=0}
\sum^1_{I=0} (2\ell+1) (2I+1)\,M^{0\,I}_{1S\,\ell}(q, q)
\,P_\ell(\cos\,\theta)\ ,\nonumber \\
& & f_3=-\frac{\mu}{2\pi \hbar^2}\,\sum^\infty_{\ell=0}
(2\ell+1)\,\left[ G^{\rm K}_\ell(q,q)+2\,\sum^1_{I=0} (2I+1)
M^{0\,I}_{1E\,\ell}(q, q)
+\varepsilon K_\ell(q,q)\right]
\nonumber \\
& & \hspace{20mm} \times P_\ell(\cos\,\theta)\ .
\label{res1}
\end{eqnarray}
Here, $\mu=(4/5)M_N$ is the reduced mass,
$q$ is the relative momentum determined
from $\varepsilon=(\hbar^2/2\mu)q^2$, and $\theta$ is the c.m.~angle
for the $n \alpha$ scattering. In \eq{res1}, the partial-wave sum
is actually taken up to $\ell=8$.
We plot $f_D+f_1$, $f_2$, and $f_3$ in Fig.\,\ref{fig7} for the
c.m.~energies $E_{\rm c.m.}=50$ and 100 MeV. For the effective local
$NN$ forces, these Born amplitudes are analytically calculated
as in Eqs.~(30) - (34) of Ref.\,\citen{TT71}. We show the results
of the MN3R force in Fig.\,\ref{fig8} for comparison. In the MN3R force,
the $f_D$ term is a Fourier transform of the momentum-independent local 
potential and $f_1$ is that of the nonlocal RGM kernel with a smaller
amplitude. Since the $f_D$ and $f_1$ terms are not separated
for the $G$-matrix interaction, the sum $f_D+f_1$ is compared
with the predictions by fss2.
The amplitudes $f_2$ and $f_3$ arise as a consequence of the 
antisymmetrization procedure and are closely related to
the nucleon-exchange effect in the backward angular region.
Figures \ref{fig7} and \ref{fig8} show that the angular
and energy dependences of these two different $NN$ interactions
are very similar to each other, in spite of the strong nonlocality
of the $G$-matrix interaction.


\section{Summary}

The $n \alpha$ system is one of the most successful examples, in which
microscopic RGM calculations using effective $NN$ forces give
a good description of the experimental data. It is, therefore, very
interesting to examine if this result is still valid when more
realistic $NN$ interactions based on bare interactions are employed
in $G$-matrix formalism. Here, we studied the $n \alpha$ system
by our previous technique \cite{B8a},
in which baryon-octet ($B_8$) $\alpha$ Born kernels are calculated
with explicit treatments of the nonlocality and the center-of-mass (c.m.)
motion between $B_8$ and $\alpha$. Since the $n \alpha$ system
involves the antisymmetrization of nucleons due to the Pauli principle,
we needed to extend the previous techniques for the direct
and knock-on terms of the hyperon $\alpha$ interaction to other
nucleon-exchange and interaction types such as the $S$ and $S^\prime$ types.
We found that the explicit treatment of the Galilean non-invariant interaction
gives some definite recoil effect to the c.m.~momentum of two interacting
nucleons, involved in the $G$-matrix, for each particular
interaction type of the RGM kernels. 
If one uses the invariant $G$-matrix as the input $NN$ interaction
for the $n \alpha$ RGM kernel, both of the direct potential and the knock-on
interaction kernel become nonlocal, and give the same contributions
for each isospin component of the $NN$ interaction with $I=0$ or 1.
This is a common feature of two-cluster systems composed of a single
nucleon and a nucleus.  

In principle, we can deal with the momentum dependence and
the starting-energy dependence of the $G$-matrix according
to the explicit expressions derived in this study.
The treatment of the Fermi momentum $k_F$ and the starting energy $\omega$,
however, needs careful treatment. Local density approximation, usually
assumed in heavier systems, cannot be justified
for $(0s)$-shell nuclei like the $\alpha$ particle.
The single-particle (s.p.)~energy used in the definition of
the starting energy is not well defined in the $n \alpha$ cluster model.
We assumed a constant $k_F$ to generate $G$-matrix $NN$ interaction
for $n \alpha$ scattering. The starting energy $\omega$ is
determined from the local momentum $q$ and the c.m.~momentum $K$
of the two interacting nucleons using angle-averaging procedure
over $\bK$ under the constraint $|\bK/2-\bq|<k_F$. The $G$-matrix calculation
is carried out in the continuous prescription
for intermediate spectra using the energy-independent version of
the quark-model $NN$ interaction fss2.
We found that, in the present framework, $k_F=1.20~\hbox{fm}^{-1}$ is
a most favorable choice for generating an appropriate strength
of the $n \alpha$ interaction. A larger $k_F$ say $k_F=1.35~\hbox{fm}^{-1}$,
gives less attractive $n \alpha$ interaction, 
and a smaller $k_F$ gives more attractive $n \alpha$ interaction.
We carried out the angular-momentum projection
of the $G$-matrix $NN$ interaction explicitly including
the $K$ dependence. Owing to this procedure, the existence
of the Pauli-forbidden $(0s)$ state for the $S$-wave relative
motion is strictly preserved, and the admixture of the redundant
component of the RGM formalism is completely eliminated.

With these treatments of $G$-matrix parameters,
we found that the central and spin-orbit components
of the $n \alpha$ Born kernel have reasonable strengths
under the assumption of a rigid translationally
invariant $(0s)^4$ shell-model wave function of the $\alpha$ cluster.
The $n \alpha$ phase shifts in the energy region, $E_n \leq 30$ MeV, are
reasonably reproduced for the $S_{1/2}$, $P_{3/2}$ and $P_{1/2}$
states, while the central attraction is somewhat too weak
for higher partial waves.
The direct comparison of the differential cross sections and
polarization with experiment shows that
these higher partial waves do not markedly impair
the fit to the experimental data, except for the energy region
where the $d+\hbox{}^3\hbox{H}$ channel opens.
In the higher energy region of up to 100 MeV in the c.m.~system,
we compared the Born amplitudes
from the $G$-matrix $NN$ interaction with those from an effective 
$NN$ force, the Minnesota three-range force. We found that
characteristic behaviors of three different groups of exchange terms,
the direct and knock-on terms, heavy-particle pickup terms,
and nucleon-rearrangement terms, are essentially the same
between these two approaches.

Note that the appropriate strength
of the $n \alpha$ central attraction in the present calculation
depends largely on how we deal with the strong starting-energy dependence
in the $G$-matrix calculation. The present procedure is one of the
possible procedures for preserving the redundancy property
of RGM formalism and still dealing with the recoil effect of the $\alpha$
cluster explicitly in the $G$-matrix $NN$ interaction.
Much simpler treatments are indeed possible, by assuming
some appropriate relative momentum $q$ and the c.m.~momentum $K$
in the starting energy $\omega(q, K)$.
We can find the most appropriate $k_F$, which strongly
correlates with the approximate reproduction of $\alpha$-cluster
internal energy. In all of such calculations, however, the strength
of the $n \alpha$ $LS$ potential is rather stable
and the necessary spin-orbit splitting between
the $P_{3/2}$ and $P_{1/2}$ resonances is always reproduced correctly.

From the present study, we learn that
our new folding procedure of the $G$-matrix elements
using simple shell-model wave functions works reasonably well
for deriving the characteristic features of the $n \alpha$ interaction.
The purpose here is not to study the validity or foundations
of the model assumptions of the five-nucleon system.
There are many unsolved conceptual problems in the realistic description
of the five-nucleon system based on the fundamental $NN$ interaction.  
Many recent studies imply the importance of the tensor correlations
and the effect of three-body forces \cite{NO07} even in the problem
of spin-orbit splitting discussed in the present paper. 
For applications of the present formalism to hyperon-nucleus potentials,
however, such details are still beyond the scope
of the present study. The basic baryon-baryon interaction itself contains
a large ambiguity, to which we hope to set some constraints
by examining the hyperon-nucleus potentials using the present approach.
Applications to $\Xi$-nucleus interactions for light nuclei,
for instance, will be discussed in the forthcoming paper. 

\bigskip

\section*{Acknowledgements}

This work was supported by Grants-in-Aids for Scientific
Research (C) (Grant Nos.~18540261 and 17540263),
and for Scientific Research on Priority
Areas (Grant No.~20028003),
and Bilateral Joint Research Projects (2006-2008)
from the Japan Society for the Promotion of Science (JSPS).
This work was also supported by the Grant-in-Aid
for the Global COE Program ``The Next Generation of Physics,
Spun from Universality and Emergence'' from the Ministry of Education,
Culture, Sports, Science and Technology (MEXT) of Japan. 

\appendix

\section{The $n \alpha$ RGM Born kernels for the Gaussian-type
effective $NN$ force}

The exchange normalization kernel
of the $n \alpha$ system in the momentum representation is given by 
\begin{eqnarray}
K(\bq_f, \bq_i) & = & f(\theta)=
\left(\frac{8\pi}{3\nu}\right)^{\frac{3}{2}}
\exp \left\{-\left(\frac{3}{32\nu}\bk^2+\frac{25}{24\nu}\bq^2\right)\right\}
\nonumber \\
& = & \left(\frac{8\pi}{3\nu}\right)^{\frac{3}{2}}
\exp \left\{-\frac{1}{3\nu}\left[
\frac{17}{16}\left({\bq_f}^2+{\bq_i}^2\right)
+ \bq_f \cdot \bq_i \right]\right\}\ ,
\label{a1}
\end{eqnarray}
where $\nu$ is the h.o.~size parameter of the $\alpha$ cluster.
In many cases, the transformation from $\bq_f$ and $\bq_i$ to
the momentum transfer $\bk$ and the local momentum $\bq$ via
$\bk=\bq_f-\bq_i$ and $\bq=\left(\bq_f+\bq_i\right)/2$ is
convenient for simplifying the expressions for the kernels.
We also use
the notation $f(\theta)$ for \eq{a1}, although it is a function
of $q_f=|\bq_f|$, $q_i=|\bq_i|$ and $\cos\,\theta=(\widehat{\bq}_f
\cdot \widehat{\bq}_i)$.
The Born kernel for the exchange kinetic-energy kernel is given by
\begin{eqnarray}
G^{\rm K}(\bq_f, \bq_i)=\frac{3\hbar^2 \nu}{2M_N}\,f(\theta)\,
\left[1-\frac{2}{3\nu}\left(\frac{5}{3}\bq^2+\frac{1}{4}\bk^2\right)\right]\ ,
\label{a2}
\end{eqnarray}
with $M_N$ being the nucleon mass.
The partial-wave decomposition of these kernels yields
\begin{eqnarray}
& & K_\ell(q_f, q_i)=(-1)^\ell
\left(\frac{8\pi}{3\nu}\right)^{\frac{3}{2}}
\exp \left\{-\frac{1}{3\nu}\left[
\frac{17}{16}\left({q_f}^2+{q_i}^2\right)-q_f q_i \right]\right\}
\nonumber \\
& & \qquad \times \widetilde{i}_\ell\left(\frac{1}{3\nu}q_f q_i\right)
\ ,\nonumber \\
& & G^{\rm K}_{\ell}(q_f, q_i)=(-1)^\ell
\left(\frac{8\pi}{3\nu}\right)^{\frac{3}{2}}
\exp \left\{-\frac{1}{3\nu}\left[
\frac{17}{16}\left({q_f}^2+{q_i}^2\right)-q_f q_i \right]\right\}
\nonumber \\
& & \qquad \times \frac{3 \hbar^2 \nu}{2M_N}
\left\{ \left[1+\frac{2}{3}\ell-\frac{4}{9\nu}\left({q_f}^2+{q_i}^2\right)
\right] \widetilde{i}_\ell\left(\frac{1}{3\nu}q_f q_i\right)
\right. \nonumber \\
& & \qquad \left. +\frac{2}{9\nu}\,q_f q_i\,\widetilde{i}_{\ell+1}
\left(\frac{1}{3\nu}q_f q_i\right) \right\}\ ,
\label{a3}
\end{eqnarray}
where $i_\lambda(x)=i^\lambda j_\lambda(-ix)
=e^x\,\widetilde{i}_\lambda(x)$.
For the Gaussian central $NN$ force in \eq{rgm7},
the spin-flavor factors in \eq{rgm6} are calculated to be 
\begin{eqnarray}
X_{0E} & = & X_d+X_e\ \ ,\qquad X_{0D_+}=\frac{1}{2} X_d\ ,\nonumber \\
X_{1E} & = & X_{1S}=X_{1S^\prime}=-\frac{1}{2}\left(X_d+X_e\right)\ \ ,
\qquad X_{1D_-}=\frac{1}{2} X_e\ ,\nonumber \\
X_d & = & 8W+4B-4H-2M\ ,\quad X_e=8M+4H-4B-2W\ .
\label{a4}
\end{eqnarray}
By using these factors, $V_{\rm D}$ and $G^{\rm V}$ are obtained as
\begin{eqnarray}
V_{\rm D}(\bq_f, \bq_i) & = & v_0 \frac{1}{2} X_d \left(\frac{\pi}{\kappa}
\right)^{\frac{3}{2}} \exp \left\{-\frac{k^2}{4}\left(\frac{3}{8\nu}
+\frac{1}{\kappa}\right)\right\}\ ,\nonumber \\
G^{\rm V}(\bq_f, \bq_i) & = & v_0 \frac{1}{2} \left(X_d+X_e\right)
\left[\,f_E(\theta)-f_S(\theta)-f_{S^\prime}(\theta)\,\right]
+ v_0 \frac{1}{2} X_e\,f_{D_-}(\theta)\ , \nonumber \\
\label{a5}
\end{eqnarray}
with
\begin{eqnarray}
& & f_\CT(\theta)=f(\theta) \nonumber \\
& & \times \left\{\begin{array}{l}
\left(\frac{1}{1+\frac{\kappa}{\nu}}\right)^{\frac{3}{2}} \\
\left(\frac{1}{1+\frac{5\kappa}{3\nu}}\right)^{\frac{3}{2}}
\exp \left\{ \frac{\frac{\kappa}{2\nu}}{1+\frac{5\kappa}{3\nu}}
\frac{1}{2\nu}\left(\frac{5}{3}\bq+\frac{1}{2}\bk\right)^2 \right\} \\
\left(\frac{1}{1+\frac{2\kappa}{\nu}}\right)^{\frac{3}{2}}
\exp \left\{ \frac{\frac{\kappa}{2\nu}}{1+\frac{2\kappa}{\nu}}
\frac{1}{2\nu} \bk^2 \right\} \\
\left(\frac{1}{1+\frac{8\kappa}{3\nu}}\right)^{\frac{3}{2}}
\exp \left\{ \frac{\frac{\kappa}{2\nu}}{1+\frac{8\kappa}{3\nu}}
\frac{50}{9\nu} \bq^2 \right\} \\
\end{array} \right.
\quad \hbox{for} \quad \CT=\left\{\begin{array}{l}
E \\[5mm]
S \\[5mm]
D_+ \\[5mm]
D_- \\[0mm]
\end{array} \right. \ .
\label{a6}
\end{eqnarray}
The $S^\prime$-type function $f_{S^\prime}(\theta)$ is obtained
from $f_S(\theta)$ by simply replacing $\bk$ by $-\bk$.
For the Gaussian $LS$ force in \eq{rgm8},
the spin-isospin factors are given by
\begin{eqnarray}
\bX^{LS}_{0D_+}=(4W-2H) \bS\ \ ,\qquad
\bX^{LS}_{1D_-}=(4H-2W) \bS\ ,
\label{a7}
\end{eqnarray}
where $\bS$ is the total spin of the $n \alpha$ system.
The $LS$ Born kernels are given by
\begin{eqnarray}
V^{LS}_{\rm D}(\bq_f, \bq_i) & = & v^{LS}_0 (4W-2H)\,\left(\frac{\pi}{\kappa}
\right)^{\frac{3}{2}} \exp \left\{-\frac{k^2}{4}\left(\frac{3}{8\nu}
+\frac{1}{\kappa}\right)\,\right\}
\frac{5}{16\kappa}\,i \bn\cdot \bS
\ ,\nonumber \\
G^{LS}(\bq_f, \bq_i) & = & v^{LS}_0 (2W-4H)\,f_{D_-}(\theta)
\left(\frac{1}{1+\frac{8\kappa}{3\nu}}\right)
\frac{5}{6\nu}\,i \bn\cdot \bS\ ,
\label{a8}
\end{eqnarray}
where $\bn=[\bq \times \bk]=[\bq_i \times \bq_f]$.
For the effective $NN$ force, the direct potentials become local
in the coordinate representation:
\begin{eqnarray}
V^{C}_{\rm D}(r) & = & v_0\,\frac{1}{2}X_d
\left(\frac{1}{1+\frac{3\kappa}{8\mu}}\right)^{\frac{3}{2}}
\exp \left\{-\frac{\kappa}{1+\frac{3\kappa}{8\nu}} r^2\right\}\ ,\nonumber \\
V^{LS}_{\rm D}(r) & = & v^{LS}_0\,(4W-2H)
\,\left(\frac{1}{1+\frac{3\kappa}{8\nu}}\right)^{\frac{5}{2}}
\frac{5}{8}\,\exp \left\{-\frac{\kappa}{1+\frac{3\kappa}{8\nu}} r^2 \right\}\ ,
\label{a9}
\end{eqnarray}
where the $LS$ potential is $V^{LS}_{\rm D}(r)\,\bfell \cdot \bS$.
The internal energy of the $\alpha$ cluster is given by
\begin{eqnarray}
E^{\rm int}_\alpha=3 \cdot \frac{3\hbar^2 \nu}{2M_N}
+v_0 \left(X_d+X_e\right) \left(\frac{\nu}{\nu+\kappa}\right)^{\frac{3}{2}}
+2 e^2 \sqrt{\frac{\nu}{\pi}}\ .
\label{a10}
\end{eqnarray}

\section{The momentum dependence of the interaction kernels
for systems of two $(0s)$-shell clusters}

In this Appendix, we will show the explicit momentum dependence
that appears in the interaction kernels of the Galilean non-invariant
two-nucleon interaction for systems of two $(0s)$-shell clusters.
For the two-nucleon interaction in \eq{int5}, the spatial part
of the interaction kernel defined in \eq{rgm6} is given by
\begin{eqnarray}
& & M_{x \CT}(\bq_f, \bq_i)=M^N_x(\bq_f, \bq_i)
\nonumber \\
& & \times \frac{1}{(2\pi)^3}
\int d \bk^\prime
\exp \left\{-\left(1+\frac{\widetilde{\alpha}}{2\mu}\right)
\frac{1}{4\nu}{\bk^\prime}^2
-\frac{1}{2\sqrt{\gamma}} \bV \cdot \bk^\prime \right\}
\nonumber \\
& & \times \left(\frac{1}{\pi\nu}\frac{1}{1-\alpha/2\mu}
\right)^{3/2} \int d \bq^\prime
\exp \left\{-\frac{1}{1-\alpha/2\mu}\frac{1}{\nu}
\left( \bq^\prime+\frac{\varepsilon}{4\mu}\bk^\prime
+\frac{\nu}{2\sqrt{\gamma}}\bA \right)^2 \right\}
\nonumber \\
& & \times
\left(\frac{A}{A-2}\frac{1}{4\pi \nu}\frac{1}{1-\beta/2\mu}\right)
^{\frac{3}{2}} \int d \bP
\exp \left\{-\frac{A}{A-2}\frac{1}{1-\beta/2\mu}\frac{1}{4\nu}
\left(\bP-\bP_0\right)^2 \right\}
\nonumber \\ [2mm]
& & \times ~u(\bk^\prime, \bq^\prime; |\bP|)\ .
\label{b1}
\end{eqnarray}
Here $M^N_x(\bq_f,\bq_i)$ is the normalization kernel given by
\begin{equation}
M^N_x(\bq_f,\bq_i)=\left(\frac{2\pi}{\gamma}
\frac{1}{1-\tau^2}\right)^{\frac{3}{2}}
\exp \left\{-\frac{1}{2\gamma}\left(\frac{1-\tau}{1+\tau}\bq^2
+\frac{1+\tau}{1-\tau}\frac{1}{4} \bk^2 \right)\right\}\ ,
\label{b2}
\end{equation}
with $\tau=1-x/\mu$ , $\bk=\bq_f-\bq_i$ and $\bq=(\bq_f+\bq_i)/2$.
We assume the mass numbers $A_1$ and $A_2$ with $A_1,~A_2 \leq 4$ for
the two clusters, $A=A_1+A_2$ is the total mass number
and $\mu=A_1 A_2/A$ is the reduced mass number. 
Almost all of the coefficients and vectors appearing in \eq{b1}
are defined in Eq.\,(A.14) of Ref.~\citen{LSRGM}.  
The $\bP$ integral is used to suitably treat the c.m.~motion of
the interacting two nucleons. 
The new parameters $\bP_0$ and $\beta$ are parametrized as
\begin{equation}
\bP_0=\frac{1}{\mu}\frac{1}{(1-\tau^2) (1-\alpha/2\mu)}\,\widetilde{\bP_0}
\ ,\quad \beta=\frac{1}{(1-\tau^2) (1-\alpha/2\mu)}\frac{A}{A-2}
\,\widetilde{\beta}\ ,
\label{b3}
\end{equation}
with $\widetilde{\bP_0}$ and $\widetilde{\beta}$ given
in Table \ref{table-a}.

\begin{table}[b]
\caption{Parameters $\widetilde{\bP_0}$ and $\widetilde{\beta}$ in
\protect\eq{b3}.}
\label{table-a}
\bigskip
\begin{center}
\renewcommand{\arraystretch}{1.4}
\setlength{\tabcolsep}{3mm}
\begin{tabular}{ccc}
\hline
$\CT$ & $\widetilde{\bP_0}$ & $\widetilde{\beta}$ \\
\hline
$E_{11}$ & $\frac{4x}{A_1}\bq$ & $8\,\frac{xA_2}{AA_1}$ \\
$E_{22}$ & $\frac{4x}{A_2}\bq$ & $8\,\frac{xA_1}{AA_2}$ \\
$E_{12}$ & $2x\left(\frac{1}{A_1}-\frac{1}{A_2}\right)\bq
+\left(2-\frac{x}{\mu}\right)\bk$ & $4\left(1-\frac{2x}{A}\right)$ \\
$E_{21}$ & $2x\left(\frac{1}{A_1}-\frac{1}{A_2}\right)\bq
-\left(2-\frac{x}{\mu}\right)\bk$ & $4\left(1-\frac{2x}{A}\right)$ \\
$S_1$ & $\left(1-\frac{2x}{A_1}\right)\left(\bq^\prime
+\frac{1}{2}\bk^\prime\right)
+\left(\frac{1-3x}{A_1}+\frac{x}{A_2}\right)\bq
+\left(2-\frac{1}{A_1}-\frac{x}{\mu}\right)\frac{1}{2}\bk$ & 1) \\
$S_2$ & $\left(1-\frac{2x}{A_2}\right)\left(\bq^\prime
+\frac{1}{2}\bk^\prime\right)
+\left(\frac{1-3x}{A_2}+\frac{x}{A_1}\right)\bq
+\left(2-\frac{1}{A_2}-\frac{x}{\mu}\right)\frac{1}{2}\bk$ & 2) \\
$S^\prime_1$ & $\left(1-\frac{2x}{A_1}\right)\left(\bq^\prime
-\frac{1}{2}\bk^\prime\right)
+\left(\frac{1-3x}{A_1}+\frac{x}{A_2}\right)\bq
-\left(2-\frac{1}{A_1}-\frac{x}{\mu}\right)\frac{1}{2}\bk$ & 1) \\
$S^\prime_2$ & $\left(1-\frac{2x}{A_2}\right)\left(\bq^\prime
-\frac{1}{2}\bk^\prime\right)
+\left(\frac{1-3x}{A_2}+\frac{x}{A_1}\right)\bq
-\left(2-\frac{1}{A_2}-\frac{x}{\mu}\right)\frac{1}{2}\bk$ & 2) \\
$D_+$ & $2x\left(\frac{1}{A_1}-\frac{1}{A_2}\right)\left(\bq^\prime+\bq\right)$
& 3) \\
$D_-$ & $(x-1)\left(\frac{1}{A_1}-\frac{1}{A_2}\right)
\left(\bk^\prime+2\bq\right)$
& 4) \\
\hline
\end{tabular}
\end{center}
\medskip
1) $1-\frac{2A_2}{AA_1}+4\frac{x}{A}\left(\frac{A_2}{A_1}-1\right)$ \\
2) $1-\frac{2A_1}{AA_2}+4\frac{x}{A}\left(\frac{A_1}{A_2}-1\right)$ \\
3) $2x\left(\frac{1}{A_1}-\frac{1}{A_2}\right)\frac{A_2-A_1}{A}$ \\
4) $2(x-1)\left(\frac{1}{A_1}-\frac{1}{A_2}\right)\frac{A_2-A_1}{A}$
\end{table}

It sometimes happens that the formula in \eq{b1} cannot be used
because of the divergence of the coefficients. For example, all the 
direct terms with $x=0$ and many of the $A_1=1$ or $A_2=1$ cases
should be treated separately.
For the direct terms, we obtain
\begin{eqnarray}
& & M_{0E_{\alpha \alpha}}(\bq_f, \bq_i)=\delta(\bk)
~\left(\frac{1}{\pi \nu}\right)^{\frac{3}{2}}
\int d \bk^\prime\,d \bq^\prime~\exp \left\{-\frac{1}{\nu}
\left({\bq^\prime}^2+\frac{1}{4}{\bk^\prime}^2 \right)\right\}
\nonumber \\
& & \times \left(\frac{1}{1-2/A_\alpha}\frac{1}{4\pi \nu}\right)^{\frac{3}{2}}
\int d \bP~\exp \left\{-\frac{1}{1-2/A_\alpha}\frac{1}{4\nu}
\left(\bP-\frac{2}{A_\alpha}\bq \right)^2\right\}
\nonumber \\
& & \times ~u(\bk^\prime, \bq^\prime; |\bP|)\ ,
\label{b4}
\end{eqnarray} 
and
\begin{eqnarray}
& & M_{0D_+}(\bq_f, \bq_i)=\exp \left\{-\left(1-\frac{1}{2\mu}\right)
\frac{1}{4\nu}\bk^2 \right\}
\left(\frac{1-1/2\mu}{(1-1/A_1)(1-1/A_2)}\frac{1}{4\pi\nu}\right)
^{\frac{3}{2}}
\nonumber \\
& & \times \left(\frac{1}{1-1/2\mu}\frac{1}{\pi\nu}\right)^{\frac{3}{2}}
\int d \bP\,d \bq^\prime
\exp \left\{-\frac{1}{1-1/2\mu}\frac{1}{\nu}\left(\bq^\prime
-\frac{1}{2\mu}\bq \right)^2 \right.
\nonumber \\
& & \left. -\frac{1-1/2\mu}{(1-1/A_1)(1-1/A_2)}\frac{1}{4\nu}
\left[\bP-\frac{1}{1-1/2\mu}\left(\frac{1}{A_1}-\frac{1}{A_2}\right)
\left(\bq^\prime-\bq\right)\right]^2 \right\}
\nonumber \\
& & \times ~u(\bk, \bq^\prime; |\bP|)\ .
\label{b5}
\end{eqnarray} 
When $A_1=1$ or $A_2=1$, \eq{b5} cannot be used either. In this case,
we obtain
\begin{eqnarray}
& & M_{0D_+}(\bq_f, \bq_i)=\exp \left\{-\left(1-\frac{1}{2\mu}\right)
\frac{1}{4\nu}\bk^2 \right\}
\left(\frac{1}{1-1/2\mu}\frac{1}{\pi\nu}\right)^{\frac{3}{2}}
\nonumber \\
& & \times \int d \bq^\prime
\exp \left\{-\frac{1}{1-1/2\mu}\frac{1}{\nu}\left(\bq^\prime
-\frac{1}{2\mu}\bq \right)^2 \right\}
~u(\bk, \bq^\prime; |2(\bq-\bq^\prime)|)
\nonumber \\
& & \hspace{60mm} \hbox{for} \quad A_1=1 \quad \hbox{or} \quad A_2=1\ .
\label{b6}
\end{eqnarray} 
The knock-on term kernel $M_{1D_-}(\bq_f, \bq_i)$ can be obtained
from Eqs.\,(\ref{b5}) and (\ref{b6}) by simply changing
$u(\bk^\prime, \bq^\prime; |\bP|)$ to $u(2\bq^\prime, \bk^\prime/2; |\bP|)$.
Finally, $1-\beta/2\mu$ becomes zero for the $S$- and $S^\prime$-type
interaction kernels in the $n \alpha$ case. In such a case,
we can use the limit formula
\begin{equation}
\lim_{\kappa \rightarrow 0} \left(\frac{1}{\kappa \pi}\right)^{\frac{3}{2}}
~e^{-{\scriptsize \bx^2}/\kappa}=\delta(\bx)\ .
\label{b7}
\end{equation}
The last two lines of \eq{b1} become
simply $u(\bk^\prime, \bq^\prime, |\bP_0|)$ using this procedure.

\end{document}